\documentclass{elsart}

\usepackage{graphicx}
\usepackage{amsfonts}
\usepackage{latexsym}
\usepackage{amssymb}
\usepackage{amsmath}

\newcommand{\ave}[1]{\left\langle #1 \right\rangle}
\newcommand{\thermave}[1]{\left\langle #1 \right\rangle_T}
\newcommand{\bigo}[1]{\mathcal{O}(#1)}
\newcommand{\binomial}[2]{{{#1} \choose {#2}}}
\newcommand{\iid}{{\it i.i.d.}}
\newcommand{\myref}[1]{(\ref{#1})}

\date{July 2000}

\begin{document}

\begin{frontmatter}
  
  \title{A physicist's approach to number partitioning}
  \author{Stephan Mertens\thanksref{EMAIL}} \address{Institut f\"ur Theoretische
    Physik, Otto--von--Guericke--Universit\"at, 39106 Magdeburg, Germany}
  \thanks[EMAIL]{http://itp.nat.uni-magdeburg.de/$\sim$mertens \\
    email: stephan.mertens@physik.uni-magdeburg.de}

\begin{abstract}
  The statistical physics approach to the number partioning problem, a classical
  NP-hard problem, is both simple and rewarding. Very basic notions and methods
  from statistical mechanics are enough to obtain analytical results for the
  phase boundary that separates the ``easy-to-solve'' from the ``hard-to-solve''
  phase of the NPP as well as for the probability distributions of the optimal
  and sub-optimal solutions.  In addition, it can be shown that solving a number
  partioning problem of size $N$ to some extent corresponds to locating the minimum in an
  unsorted list of $\bigo{2^N}$ numbers.  Considering this correspondence it is not
  surprising that known heuristics for the partitioning problem
  are not significantly better than simple random search.
\end{abstract}


\end{frontmatter}

\section{Introduction}
\label{sec:intro}

Recent years have witnessed an increasing interaction among the disciplines of
discrete mathematics, computer science, and statistical physics. These fields
are linked by the fact that models from statistical physics can be formalized as
combinatorial optimization problems and vice versa
\cite{mezard:etal:87,rieger:98}. The connection between optimization and
statistical physics has lead to practical algorithms like simulated annealing
\cite{kirkpatrick:gelatt:vecchi:83} and to new theoretical results, some of
which can be found in this special issue.

In most cases where a statistical physics analysis of an optimization or
decision problem yields significant new results, this analysis is rather
complicated, technically as well as conceptionally. This complexity may easily
deter computer science people from learning the tricks and tools, even if they
value the results. To promote interdisciplinarity beyond the mutual
appreciation of results, it may help to consider a physicists approach to an
optimization problem, which on the one hand requires only very basic notions and
methods from statistical mechanics, but on the other hand yields non trivial
results. In fact there exists a problem with this nice property: the {\em number
  partitioning problem}.  It is defined as follows: Given a list
$a_1,a_2,\ldots,a_N$ of positive numbers, find a partition, i.e.\ a subset
$\mathcal{A}\subset\{1,\ldots,N\}$ such that the {\em partition difference}
\begin{equation}
  \label{eq:cost-function}
  E(\mathcal{A}) = \Big|\sum_{i\in \mathcal{A}}a_i - \sum_{i\not\in \mathcal{A}} a_i\Big|,
\end{equation}
is minimized.  In the constrained partition problem, the cardinality difference
between $\mathcal{A}$ and its complement,
\begin{equation}
  \label{eq:magnetization}
  M = |\mathcal{A}|-(N - |\mathcal{A}|) = 2|\mathcal{A}| - N,
\end{equation}
is fixed. A special case is the {\em balanced partitioning problem} with the
constraint $|M| \leq 1$.

Partitioning is of both theoretical and practical importance. It is one of Garey
and Johnson's six basic NP-complete problems that lie at the heart of the theory
of NP-completeness \cite{garey:johnson:79}.  Among the many practical
applications one finds multiprocessor scheduling and the minimization of VLSI
circuit size and delay \cite{coffman:lueker:91,tsai:92}.

In this paper we present a statistical mechanics approach to the NPP.  We start,
however, with a brief discussion of some known facts about the NPP.  We will
learn that there is a {\em phase transition} in the computational complexity of
the NPP, and that there are {\em no really good heuristics} for this problem.
Both facts will be discussed within the framework of statistical mechanics in
the following sections. Section \ref{sec:statmech} starts with an introduction
into the very basic notions and methods of statistical mechanics. We formulate
the NPP as a spin-glass, i.e.\ as a model to describe magnetic alloys, and
calculate its free energy and entropy. The entropy in turn yields a simple
analytic expression for the phase boundary that separates the ``easy-to-solve''
from the ``hard-to-solve'' phase in the NPP. In addition, we get an expression
for the average optimum partition difference. The statistical
mechanics analysis reveals another phase transition in the constrained NPP: if
$M$ exceeds a critical value, the NPP becomes overconstrained and its solution
trivial.  In section \ref{sec:random} we map the balanced and the
unconstrained NPP to another physical model, the
random energy model. This signifies that solving the NPP with $N$ random numbers
$a_j$ corresponds to locating the minimum in an unsorted list of $\bigo{2^N}$
random numbers. This correspondence provides us with an explanation of the bad performance
of heuristic algorithms for the NPP and in addition allows us to derive 
analytical expressions for the {\em probability
  distribution} of the optimal and sub-optimal costs.

%%% Local Variables: 
%%% mode: latex
%%% TeX-master: "main"
%%% End: 
 \section{Some facts about number partitioning}
\label{sec:simple}

The computational complexity of the number partitioning problem depends on the
type of input numbers $\{a_1,a_2,\ldots,a_N\}$.  Consider the case that the
$a_j$'s are positive integers bound by a constant $A$. Then the cost $E$ can
take on at most $NA$ different values, i.e.\ the size of the search space is
$\bigo{NA}$ instead of $\bigo{2^N}$ and it is very easy to devise an algorithm
that explores this reduced search space in time polynomial in $NA$.
Unfortunately, such an algorithm does not prove $P=NP$ since a concise encoding
of an instance requires $\bigo{N\log A}$ bits, and $A$ is not bounded by any
polynomial of $\log A$. This feature of the NPP is called ``pseudo polynomiality''.
The NP-hardness of the NPP requires input numbers of arbitrary size or, after division by
the maximal input number, of unlimited precision.

To study typical properties of the NPP, the input numbers are usually taken to
be independently and identically distributed (\iid) random numbers, drawn from
``well behaved'' distributions. Under this probabilistic assumption, the minimal
partition difference $E_1$ is a stochastic variable. For real valued input
numbers (infinite precision, see above), Karmarkar, Karp, Lueker and Odlyzko
\cite{karmarkar:etal:86} have proven that the {\em median} value of $E_1$ is
$\bigo{\sqrt{N}\cdot2^{-N}}$ for the unconstrained and $\bigo{N\cdot 2^{-N}}$
for the balanced NPP. Lueker \cite{lueker:98} showed recently, that the same
results hold for the {\em average} value of $E_1$.  Numerical simulations
\cite{ferreira:fontanari:98} indicate, that the relative width of the
distribution of $E_1$, defined as
\begin{equation}
  \label{eq:def-r}
  r := \frac{\sqrt{\ave{E_1^2} - \ave{E_1}^2}}{\ave{E_1}},
\end{equation}
where $\ave{\cdot}$ denotes the average over the $a_j$'s, tends to $1$ in the
limit $N\to\infty$, for both the unconstrained and the balanced partitioning
problem. This means, that the typical fluctuations of $E_1$ are of the same size
than the value itself.  In section \ref{sec:random} we will calculate the
complete probablity distribution of $E_1$ and rederive all these results.

Another surprising feature of the NPP is the {\em poor performance of heuristic
  algorithms} \cite{johnson:etal:91,ruml:etal:96}. In section
\ref{sec:random} we show that the bad efficiency of heuristics approaches can be
understood by the observation that number partitioning is essentially equivalent
to locating the minimum in an unsorted list of $\bigo{2^N}$ random numbers
\cite{mertens:00a}. Here we will describe some of the heuristics.

The key ingredient to the most powerful partition heuristics is the differencing
operation \cite{karmarkar:karp:82}: select two elements $a_i$ and $a_j$ and
replace them by the element $|a_i-a_j|$. Replacing $a_i$ and $a_j$ by
$|a_i-a_j|$ is equivalent to making the decision that they will go into opposite
subsets. Applying differencing operations $N-1$ times produces in effect a
partition of the set $\{a_1,\ldots,a_N\}$. The value of its partition difference
is equal to the single element left in the list.  Various partitions can be
obtained by choosing different methods for selecting the pairs of elements to
operate on. In the {\em paired differencing method} (PDM), the elements are
ordered.  The first $\lfloor N/2\rfloor$ operations are performed on the largest
two elements, the third and the fourth largest, etc..  After these operations,
the left-over $\lceil N/2 \rceil$ elements are ordered and the procedure is
iterated until there is only one element left.  Another example is the
Karmarkar-Karp (KK) or {\em
  largest differencing method} \cite{karmarkar:karp:82}. 
Again the elements are ordered.  The
largest two elements are picked for differencing. The resulting set is ordered
and the algorithm is iterated until there is only one element left.  The time
complexity of PDM and KK is $\bigo{N\log N}$, the space-complexity is
$\bigo{N}$.

\begin{figure}[htbp]
  \includegraphics[width=\columnwidth]{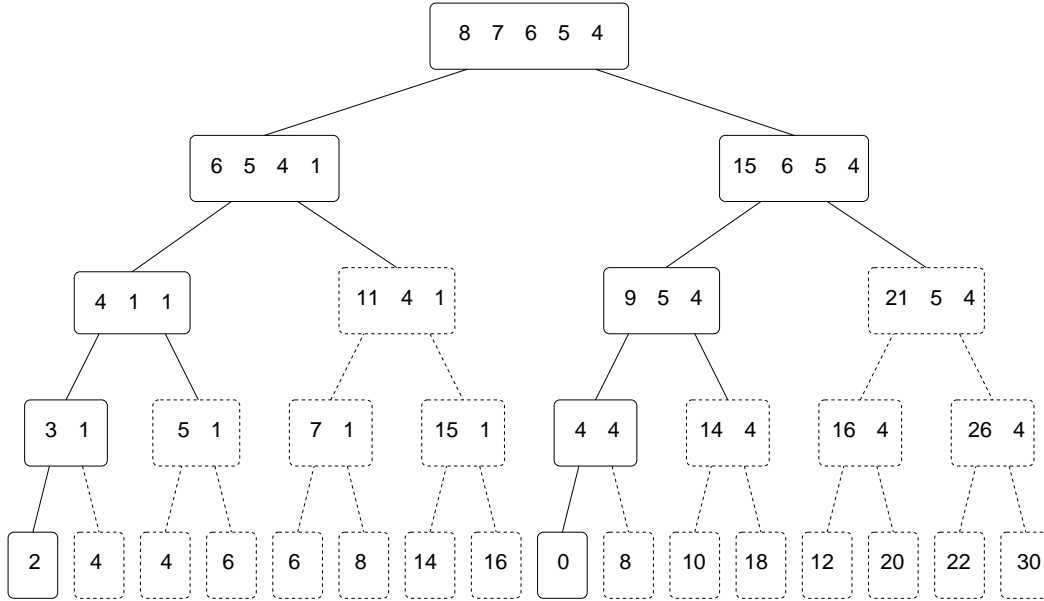}
  \caption{\small Tree generated by the CKK algorithm on the list $8,7,6,5,4$.
    Left branch: Replace the two largest numbers by their difference.  Right
    branch: Replace the two largest numbers by their sum.  The dashed parts of
    the tree are pruned by the algorithm. Thanks to the pruning rules, only 9 of
    31 nodes have to be explored.}
  \label{fig:example}
\end{figure}

The Karmarkar-Karp differencing is the best known heuristics for the partioning problem, 
but it finds an
approximate solution only, far away from the true optimum. 
KK  yields unconstrained
partitions with expected difference $\bigo{N^{-a\log N}}$, which has to be 
compared to $\bigo{\sqrt{N}\cdot 2^{-N}}$ for the true optimum. 
Korf \cite{korf:98}
showed, how the KK differencing can be extended to a {\em complete anytime algorithm}, i.e.\ 
an algorithm that finds better and better solutions the longer it is allowed to
run, until it finally finds and proves the optimum solution: At each iteration,
the KK heuristic commits to placing the two largest numbers in different
subsets, by replacing them with their difference.  The only other option is to
place them in the same subset, replacing them by their sum. This results in a
binary tree, where each node replaces the two largest remaining numbers,
$a_1\geq a_2$: the left branch replaces them by their difference, while the
right branch replaces them by their sum:
%\begin{equation}
%  \begin{array}{ccccc}
%    & & x_1, x_2, x_3, \ldots & & \\
%    & \swarrow & & \searrow & \\
%    x_1-x_2, x_3, \ldots & & & & x_1+x_2, x_3, \ldots
%  \end{array}
%\end{equation}
\begin{equation}
  a_1, a_2, a_3, \ldots \mapsto \left\{
  \begin{array}{rl}
    |a_1-a_2|, a_3, \ldots & \mbox{ left branch } \\
    a_1+a_2, a_3, \ldots & \mbox{ right branch } 
  \end{array}
  \right.
\end{equation}
Iterating both operations $N-1$ times generates a tree with $2^{N-1}$ terminal
nodes.  The terminal nodes are single element lists, whose elements are the
valid partition differences.  Korf's complete Karmarkar-Karp differencing algorithm 
(CKK) searches this
tree depth-first and from left to right. The algorithm first returns the KK-heuristic
solution, then continues to find better solutions as time allows.  See
Fig.~\myref{fig:example} for the example of a tree generated by the CKK.

There are two ways to prune the tree: At any node, where the difference between
the largest element in the list and the sum of all other elements is larger than
the current minimum partition difference, the node's offspring can be ignored.
For integer valued $a_j$, a partition with $E \leq 1$ is called {\em perfect}.
If one reaches a terminal node with a perfect partition, the entire search can
be terminated since no improvement is possible.  The dashed nodes in
Fig.~\myref{fig:example} are pruned by these rules.

\begin{figure}[htbp]
  \includegraphics[width=\columnwidth]{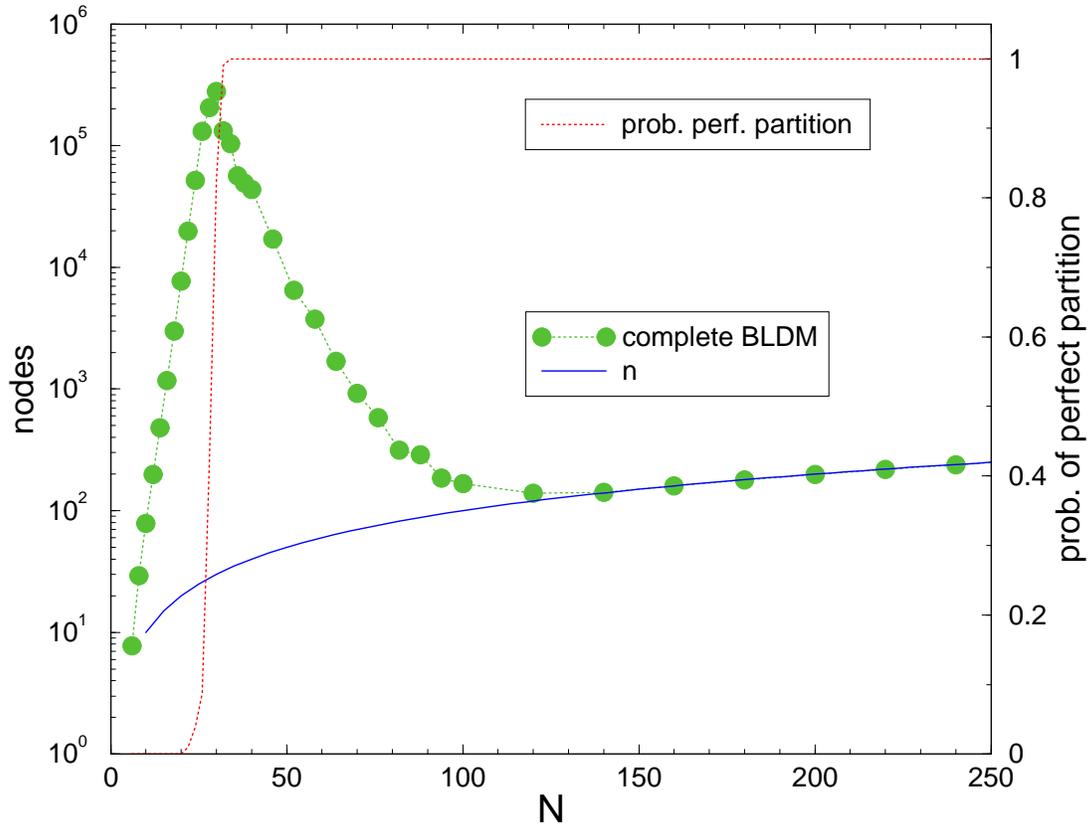}
  \caption{\small Number of nodes generated by the complete BLDM algorithm to optimally 
    partition random 25-bit integers. The complete BLDM algorithm is a variant
    of Korf's complete differencing method, modified to solve the balanced NPP
    \cite{mertens:99a}.}
  \label{fig:korf}
\end{figure}

A variant of Korf's complete differencing method to solve the balanced NPP is
called the complete balanced largest differencing method, BLDM
\cite{mertens:99a}.  It works very similar to the Korf-algorithm but generates
only balanced partitions.  To measure the performance of this algorithm as an
exact solver, we count the number of nodes generated until the optimum solution
has been found and proven.  The result for $25$-bit integers is shown in
Fig.~\myref{fig:korf}.  Each data point represents the average of $100$ random problem
instances. The horizontal axes shows the number of integers to be partitioned, the
vertical axes shows the number of nodes generated (left) and the fraction of
instances that have a perfect partition (right).  Note that we counted all nodes
of the tree, not just the terminal nodes.  We observe three distinct regimes:
for $N < 30$, the number of nodes grows exponentially with $N$, for $N > 30$ it
{\it decreases} with increasing $N$, reaching a minimum for $N\approx 130$ and
starting to grow like $N$ for larger values of $N$.

The region of exponential growth is characterized by the lack of perfect
partitions.  In this regime, the algorithm has to search the whole tree in order
to find and prove the optimum partition. For $N > 30$ it finds a perfect
partition and stops the search prematurely. The number of perfect partitions
seems to increase with increasing $N$, making it easier to find one of them.
This would explain the decrease of search costs.  For $N \gg 30$, the KK
partition already is perfect: The construction of this first partition always
requires $N$ nodes ($N-1$ internal nodes and one leaf).

Numerical simulations show that this a general scenario in the NPP.  For \iid{}
random $b$-bit numbers $a_j$, the solution time grows exponentially with $N$ for
$N \lesssim b$ and polynomially for $N \gg b$
\cite{gent:walsh:96,korf:95,korf:98}.  Problems with $N\sim b$ require the
longest solution time.  The transition from the ``hard'' to the computational
``easy'' phase has some features of a phase transition in physical systems.
Phase transitions of this kind have been observed in numerous NP-complete
problems \cite{cheeseman:etal:91,gent:walsh:95,monasson:etal:99}, and can often
be analyzed quantitatively in the framework of statistical mechanics. Compared
to other problems, this analysis is surprisingly simple for the number
partitioning problem \cite{mertens:98a}.  This is what we will do in the next
section.

%%% Local Variables: 
%%% mode: latex
%%% TeX-master: "main"
%%% End: 
 \section{Phase transitions}
\label{sec:statmech}

The observed transition from a computationally ``hard'' to an ``easy'' regime in the
NPP can well be analysed within the framework of statistical mechanics.
We start with a very brief and superficial sketch of statistical mechanics and how it can
be used to study combinatorial optimization, introducing the basic
quantities {\em free energy} and {\em entropy} and rewriting the NPP as
a physical model system, a {\em spin glass}. 
Then we calculate the free energy and the entropy of the unconstrained NPP,
learning some tools from the physicists toolbox like the $\delta$-function and 
the Laplace method to evaluate integrals.
The entropy allows us to define a precice expression for the {\em control parameter}
that fixes whether the NPP is ``hard'' or ``easy''. After that we try the same
calculation for the constrained NPP, which is a bit more cumbersome. It turns
out that a complete solution of the general constrained NPP requires some numerics,
but the balanced NPP can be solved analytically. Again we find the control parameter and
its critical value. The numerical solution of the general constrained NPP reveals the existence
of another phasetransition from a computational hard to computational easy phase. 
The control parameter for this phase transition is $M$, the imposed cardinality difference. 
This phase transition is dicussed in the last section.

\subsection{Statistical mechanics, optimization and spin glasses}
\label{sec:spin-glasses}

The aim of statistical mechanics is to predict the properties of systems
composed of very large numbers of particles in terms of the mechanical
properties of the individual particles and of the forces between them.  How
large is ``very large''? A few gramm of matter consists of about $10^{23}$
atoms. The state of a system is specified by the position, velocity,
magnetization, \ldots of each of these atoms. The equations that describe the
evolution of this {\em microstate} in time are known in princible, but it is
completely hopeless to solve them for $10^{23}$ particles.  From everyday
experience we know however, that the temperature and the pressure of a gas in a
vessel do not change in time, although the microstate, i.e.\ the positions and
velocities of all the gas atoms, keeps changing all the time.  Hence the
macroscopic properties of a system are not sensitive to its particular
microscopic state.  This physical variant of the law of large numbers
constitutes the starting point of statistical mechanics: Instead of determining
the exact value of a macroscopic quantity in a single system, its {\em average}
value is computed, taken over a suitable ensemble of similarly prepared systems.
The ensemble average value is usually much easier to compute than the exact
value. Note that this works {\em because} of the large number $N$ of particles.
The results of statistical mechanics are valid only in the $N\to\infty$, the so called
thermodynamic limit. The use of a capital letter $N$ shall remind you of this.

Consider a system with possible microstates $s\in{\mathcal S}$. For a gas, $s$
contains the positions and velocities of all atoms. If the system is kept at a
temperature $T$, according to statistical mechanics macroscopic quantities (like
the pressure) can be calculated as averages over the {\em canonical ensemble},
in which each microstate has a probability
\begin{equation}
  \label{eq:def-gibbs-prob}
  p(s) = \frac{1}{Z} e^{-H(s)/T}.
\end{equation}
$H(s)$ is termed the Hamiltonian of the system. It is a real valued function
that yields the {\em energy} of the microstate $s$. The normalization factor
\begin{equation}
  \label{eq:def-Z}
  Z = \sum_{s\in{\mathcal S}}e^{-H(s)/T}.
\end{equation}
is called the {\em partition function}, not to be mixed up with the partition in the
NPP. The {\em thermal average} of a quantity $A$ is given by
\begin{equation}
  \label{eq:def-thermal-average}
  \thermave{A} = \frac{1}{Z}\sum_{s\in{\mathcal S}} A(s) e^{-H(s)/T}.
\end{equation}
The thermal average no longer depends on a particular microstate, but only on
macroscopic parameters like the temperature $T$, reflecting precisely the
experimental observations.

The central quantity that is calculated in statistical mechanics, is the {\em
  free energy} $F$,
\begin{equation}
  \label{eq:def-F}
  F(T) = -T\ln Z.
\end{equation}
Once $F$ is known as a function of the temperature $T$ and other relevant
parameters like volume or magnetic field, most properties of the system can
easily be calculated. At least this is what they tell you in textbooks on
statistical physics. The thermal average of the energy for example is given by
\begin{equation}
  \label{eq:thermave-energy}
  \thermave{H} = \frac{1}{Z}\sum_{s\in{\mathcal S}} H(s) e^{-H(s)/T} 
               = T^2 \frac{\partial}{\partial T} F(T).
\end{equation}

What has all this to do with combinatorial optimization? Well, we can formally
define a free energy for any optimization problem: ${\mathcal S}$ is the set of
all feasible solutions, and $H(s)$ is the cost function that has to be
minimized.  This free energy comprises some usefull information about the
optimization problem.  Let $E_1 < E_2 < E_3 < \ldots$ be the sorted list of
possible values of the cost function and $n(E_k)$ be the number of feasable
solutions that yield $E=E_k$. Then the free energy is
\begin{eqnarray*}
  F(T) &=& -T\ln Z \\
       &=& -T\ln \sum_{k=1} n(E_k)\,e^{-E_k/T} \\
       &=& -T\ln \left[ n(E_1) e^{-E_1/T} (1 + \frac{n(E_2)}{n(E_1)}
           e^{-\frac{E_2-E_1}T} + \ldots )\right] \\
       &=& E_1-T\ln n(E_1)-\ln(1 + \frac{n(E_2)}{n(E_1)}
           e^{-\frac{E_2-E_1}T} + \ldots)
\end{eqnarray*}
From that we get the {\em value} of the optimum solution
\begin{displaymath}
 \lim_{T\to 0} F(T) =  E_1
\end{displaymath}
as well as the logarithm of the {\em number} of optimum solutions
\begin{displaymath}
  \lim_{T\to 0} -\frac \partial{\partial T} F(T) = \lim_{T\to 0} S(T) = \ln n(E_1).
\end{displaymath}
In physics jargon, $S(T)$ is the {\em entropy}, $E_1$ the {\em ground state
  energy}.  By adding additional terms to the cost function and recalculating
the free energy, more informations can be obtained, for instance on the
structure of the optimum solution. If you do not like all the physics jargon you 
might consider the free energy as a kind of {\em generating function} that encodes
properties of your combinatorial optimization problem.

A class of models that have been intensely investigated in physics are {\em spin
  glasses} \cite{mezard:etal:87}. In its simplest form, the microstate of a spin
glass is a set of $N$ binary variables, $s_j=\pm 1$, $j=1,\ldots,N$, called
Ising spins.  With spin glass models, physicists try to capture the properties
of magnetic alloys.  An Ising spin is the magnetic moment of an atom that can
only be oriented along a given axis in space, either ``up'' ($s_j=+1$) or
``down'' ($s_j=-1$). In an alloy these moments interact, giving rise to a total
energy
\begin{equation}
  \label{eq:spin-glass-hamiltonian}
  H(\{s_j\}) = -\sum_{i,j=1}^N J_{ij}s_is_j.
\end{equation}
The $J_{ij}$ are numbers that describe the interaction strength between spins
$s_i$ and $s_j$. The calculation of the interactions $J_{ij}$ (as well as the
justification for the whole model) is a subject of quantum mechanics and will
not be discussed here.

Minimizing the spin glass hamiltonian for given interactions $J_{ij}$ is a
combinatorial optimization problem. If all the $J_{ij}$ are positive (physics
jargon: ferromagnetic), this problem is trivial: $H$ is minimized when all spins
point in the same direction, i.e.\ are all $+1$ or all $-1$.  If some (or all)
of the $J_{ij}$ are negative (physics jargon: anti-ferromagnetic), this problem
is much harder. In fact it can be proven that it is NP-hard \cite{barahona:82}.

A partition ${\mathcal A}$ in the number partitioning problem can be encoded by
Ising spins: $s_j=+1$ if $j\in{\mathcal A}$, $s_j=-1$ otherwise. The cost
function then reads
\begin{equation}
  \label{eq:costfunction_spins}
  E = |\sum_{j=1}^Na_js_j|,
\end{equation}
and the minimum partition is equivalent to the ground state of the Hamiltonian
\begin{equation}
  \label{eq:hamiltonian}
  H = E^2 = \sum_{i,j=1}^N s_i\,a_ia_j\,s_j.
\end{equation}
This is an infinite range Ising spin glass with antiferromagnetic couplings
$J_{ij}=-a_ia_j < 0$.  The statistical mechanics of this model has been
investigated in physics at least three times
\cite{fu:89,ferreira:fontanari:98,mertens:98a}.  It turns out that due to the
multiplicative character of the couplings, the calculation of the free energy is
comparatively simple, and yields quantitative results on the phase transition in
computational complexity \cite{mertens:98a}.

Of course we are not interested in a particular instance but in the {\em
  typical} properties of number partitioning. Hence we will average our results
over a suitable {\em ensemble of instances}, not to be mixed up with the
thermodynamic ensemble of microstates resp.\ feasible solutions. Throughout this
paper we will assume that the input numbers $a_j$ are independent, identically
distributed (\iid) random numbers.  In our statistical mechanics framework,
random input numbers correspond to random spin interactions $J_{ij}$. In fact
this is part of the definition of spin glass models -- the term ``glass'' refers to the
irregularity of the interactions in alloys as opposed to regular interactions in
crystalls. In spin glass theory, it is the free energy that has to be averaged
over the random couplings to yield the correct typical properties of the system.
In general, the computation of the average free energy is not simple and requires
a sophisticated approach called the replica method. The free energy
of the number partitiong problem is so simple though, that we get its
average for free. 

\subsection{Statistical mechanics of the unconstrained NPP}
\label{sec:unconstrained-NPP}

We start with the statistical mechanics of the unconstrained NPP. This analysis
has been published elsewhere \cite{mertens:98a}, but the presentation here is
more comprehensive.  The partition function of the unconstrained NPP reads
\begin{equation}
  \label{eq:Z-unconstrained-1}
  Z = \sum_{\{s_j\}} e^{-\frac1T|\sum_ja_js_j|}.
\end{equation}
Without the absolute value in the exponent, this sum can easily be calculated:
\begin{eqnarray}
  \label{eq:easy-sum}
  \sum_{\{s_j\}} e^{-\frac{1}T\sum_ja_js_j}
     &=& \sum_{\{s_j\}} \prod_{j=1}^Ne^{-\frac{1}T a_js_j}\nonumber\\
     &=& \sum_{s_1=\pm1}e^{-\frac{1}T a_1s_1}\cdot\sum_{s_2=\pm1}
          e^{-\frac{1}T a_2s_2}\cdot \ldots
         \cdot \sum_{s_N=\pm1}e^{-\frac{1}T a_Ns_N}\nonumber\\
     &=& 2\cosh{\frac{a_1}{T}}\cdot 2\cosh{\frac{a_2}{T}}\cdot \ldots 
           \cdot 2\cosh{\frac{a_N}{T}}\nonumber\\
     &=& 2^N\prod_{j=1}^N\cosh{\frac{a_j}T}
\end{eqnarray}
The question is, how can we get rid of the absolute value in the exponent?  A
standard trick in statistical mechanics to remove nasty nonlinearities like this
is the creative use of the $\delta$-function. Introduced by P.A.M.~Dirac on an
intuitive base in connection with quantum mechanics, it is now embedded in an
exact mathemetical framework \cite{lighthill:59}. Here we stick to the more
intuitive picture and define the $\delta$-function via its Fourier integral,
\begin{equation}
  \label{eq:dirac-delta-fourier}
  \delta(x) = \frac{1}{2\pi}\int_{-\infty}^{\infty} d\hat{x} e^{i x\hat{x}}.
\end{equation}
$\delta(x)$ is $0$ for $x\neq 0$ and $\infty$ for $x=0$, and the peak at $x=0$
is perfectly calibrated to give
\begin{equation}
  \label{eq:dirac-delta-integral}
  \int_{-\infty}^\infty dx f(x)\delta(x-c) = f(c)
\end{equation}
for any reasonably well behaved function $f$.  The $\delta$-function helps us to
separate the absolute value from the summation variables $s_j$:
\begin{eqnarray*}
  \label{eq:Z-unconstrained-2}
  Z &=& \sum_{\{s_j\}} \int_{-\infty}^{\infty}
  \!\!dx \,e^{-|x|}\, \delta(x - \frac{1}{T}\sum_{j=1}^N{a_js_j}) \\
    &=& \int_{-\infty}^{\infty}
  \!\!dx \,e^{-|x|}\, \frac{1}{2\pi}\int_{-\infty}^{\infty} d\hat{x} 
     e^{i x\hat{x}} \sum_{\{s_j\}} 
  e^{-i\frac{\hat{x}}{T}\sum_ja_js_j}
\end{eqnarray*}
Now we can carry out the summation over the $\{s_j\}$ like in
eq.~\myref{eq:easy-sum}:
\begin{equation}
  \label{eq:Z-unconstrained-3}
  Z = 2^N \int_{-\infty}^{\infty}\frac{d\hat{x}}{2\pi}\prod_{j=1}^N\cos(
  \frac{a_j}{T}\hat{x})\int_{-\infty}^{\infty}\!\!dx\,e^{-|x|+i\hat{x}x}.
\end{equation}
Note that $\cosh(i x) = \cos(x)$. Doing the $x$-integral,
\begin{equation}
  \label{eq:x-integration}
  \int_{-\infty}^{\infty}\!\!dx\,e^{-|x|+i\hat{x}x} = \frac{2}{1+\hat{x}^2},
\end{equation}
and substituting $y = \arctan\hat{x}$ finally leads us to
\begin{equation}
  \label{eq:Z-unconstrained-4}
  Z = 2^N \int_{-\frac{\pi}2}^{\frac{\pi}2}\!\!\frac{dy}\pi e^{N\,G(y)}
\end{equation}
with
\begin{equation}
  \label{eq:G-unconstrained-exact}
  G(y) = \frac1N\sum_{j=1}^N\ln{\cos(\frac{a_j}{T}\tan(y))}.
\end{equation}

For large values of $N$, the statistical independence of the $a_j$ allows us to
apply the law of large numbers, i.e.\ to replace the sum by the average over
$a$:
\begin{equation}
  \label{eq:G-unconstrained}
  G(y) \approx \ave{\ln{\cos(\frac{a}{T}\tan(y))}}.
\end{equation}
This replacement is the main reason why spin glasses with couplings that factorize,
$J_{ij} = -a_ia_j$, are
comparatively easy to solve \cite{provost:vallee:83}. It relieves us from
averaging $\ln Z$, which can be very difficult in other spin glass models.

The integral in eq.~\myref{eq:Z-unconstrained-3} can be evaluated asymptotically
for large $N$ using the {\em Laplace method}: The general idea is, that the
integral is dominated by the contributions from the maxima of $G(y)$. If $G(y)$
has a maximum at $y=y_0$,
\begin{equation}
  \label{eq:laplace-example}
  \int e^{N G(y)} dx \approx e^{N G(y_0)} \int e^{-\frac{N}{2}G''(y_0)(y-y_0)^2} dy =
  e^{N G(y_0)} \sqrt{\frac{2\pi}{N G''(y_0)}}
\end{equation}
for large $N$. A general discussion of the Laplace method for the asymptotic
expansion of integrals can be found in various text-books
\cite{debruijn:61,mathews:walker:book}.

To find the maxima of $G(y)$, we will assume that $a$ can only take on
values that are integer multiples of a fixed number $\Delta a$. For integer
distributions $\Delta a = 1$, and for floatingpoint distributions $\Delta a$ is
the smallest number that can be represented with the available number of bits.
This is a reasonable assumption since we know, that the properties of the NPP
depend on the resolution in $a$. With this assumption, the solutions of
\begin{equation}
  \label{eq:dG-dY-unconstrained}
  G'(y) = \ave{\frac{a}{T}\tan\big(\frac{a}{T}\tan y\big)} \cdot (1+\tan^2 y) = 0
\end{equation}
are given by
\begin{equation}
  \label{eq:saddlepoints-unconstrained}
  y_k = \arctan\Big(\frac{\pi T}{\Delta a} k\Big) \qquad k = 0, \pm1, \pm2, \ldots.
\end{equation}
Note that $\tan\big(\frac{a}{T}\tan y_k\big) = 0$ for all values $a=n\cdot\Delta
a$.  Of course we have to consider the contributions of all saddle points when
evaluating the integral in Eq.~\myref{eq:Z-unconstrained-3}:
\begin{equation}
  Z \approx 2^N \sum_k \int_{-\infty}^{\infty} \frac{dy}{\pi} 
     e^{-\frac{N}{2}G''(y_k)y^2} 
          = 2^N \frac{\sqrt{2}}{\sqrt{\pi N}} \sum_k \frac{1}{\sqrt{G''(y_k)}}.
\end{equation}
With
\begin{equation}
  \label{eq:dG-dydy-unconstrained}
  G''(y_k) = \frac{\ave{a^2}}{T^2} \left[1+\left(\frac{\pi T}
  {\Delta a}\right)^2 k^2\right]^2
\end{equation}
and the useful identity
\begin{equation}
  \label{eq:coth-series-1}
  \sum_{k=0,\pm1,\ldots} \frac{1}{1+(xk)^2} = \frac{\pi}{x} \cdot \coth\frac{\pi}{x}
\end{equation}
we finally get
\begin{equation}
  \label{eq:Z-unconstrained}
  Z = 2^N \cdot \frac{\Delta a}{\sqrt{\frac{\pi}{2} N \ave{a^2}}} \cdot \coth
  \frac{\Delta a}{T}.
\end{equation}

The partition function $Z$ immediately yields the free energy
\begin{equation}
  \label{eq:F-unconstrained}
  F(T) = -T N \ln2 + \frac{T}{2}\ln\frac{\pi N \ave{a^2}}{2 \Delta a^2} - 
  T \ln\coth\frac{\Delta a}{T}
\end{equation}
and the thermal average of the energy
\begin{equation}
  \label{eq:thermE-unconstrained}
  \thermave{E} = \frac{\Delta a}{\sinh\frac{\Delta a}{T} \cosh\frac{\Delta a}{T}}.
\end{equation}
Let $\Delta a > 0$ be fixed. Then $\lim_{T\to 0}\thermave{E} = 0$, i.e.\ the
ground states are perfect partitions. How many perfect partions can we expect?
The answer is given by the entropy $S$, which according to
eq.~\myref{eq:Z-unconstrained} and eq.~\myref{eq:def-F} can be written as
\begin{equation}
  \label{eq:entropy}
  S = N(\kappa_c-\kappa)\ln2
      + \tilde{S}(\frac{\Delta a}{2T}),
\end{equation}
with
\begin{equation}
  \label{eq:kappa_c}
  \kappa_c = 1 - \frac{\ln\big(\frac{\pi}{6} N\big)}{N\,2\ln2}
\end{equation}
 \begin{equation}
  \label{eq:kappa}
  \kappa = \frac{\ln \frac{3}{\Delta a^2}\ave{a^2}}{N\,2\ln2},
\end{equation}
and the thermal contribution to the entropy is
\begin{equation}
  \label{eq:thermal_entropy}
  \tilde{S}(\frac{\Delta a}{T}) = \ln\coth\frac{\Delta a}{T} 
  + \frac{\Delta a}{T} \, \frac{\coth^2\frac{\Delta a}{T}-1}
  {\coth\frac{\Delta a}{T}}.
\end{equation}
For finite $\Delta a$, $\tilde{S}$ vanishes at zero temperature and increases
monotonically with $T$. In this case, the zero temperatur entropy is given by
$N(\kappa_c-\kappa)\ln2$.  If $\kappa < \kappa_c$, we have an extensive entropy
resp.\ an exponential number of perfect partitions.  $N\cdot\kappa$ is a measure
for the number of bits needed to encode the $a_j$'s.  Let the $a_j$ be \iid{}
$b$-bit integer numbers. Then $\Delta a = 1$ and
\begin{equation}
  \label{eq:kappa-special}
  \kappa = \frac{b}{N} + \frac{1}{2N}\ln_2\big(1-\frac{3}{2}2^{-b}+
    \frac{1}{2}2^{-2b}\big) =
           \frac{b}{N} + \frac{1}{N}\cdot\bigo{2^{-b}}.
\end{equation}
In this case the condition $\kappa < \kappa_c$ translates into
\begin{equation}
  \label{eq:phase-boundary-special}
  b < N - \frac{1}{2}\ln_2\big(\frac{\pi}{6}N\big).
\end{equation}
This inequality must be fullfilled in order to have perfect partitions.  The
first term on the right hand side can be explained within a simple
approximation \cite{gent:walsh:96}: Let the $N$ numbers $a_i$ each be
represented by $b$ bits. Now consider the partition difference $E$ {\em
  bitwise}. About half of all partitions will set the most significant bit of
$E$ to zero. Among those partitions, about one half will set the second most
significant bit to zero, too. Repeating this procedure we can set at most $N$
bits to zero until running out of available partitions.  To get a perfect
partition with all $b$ bits being zero, $N$ must be larger than $b$.  
This consideration ignores the carry
bits, which lead to the logarithmic
corrections in eq.~\myref{eq:phase-boundary-special}.

What happens if $N < b$ resp.\ $\kappa > \kappa_c$?  According to the
approximative consideration above we expect the optimum partition difference to
be exponentially small, $\bigo{2^{-N}}$, but larger than zero.  It looks as if
the zero temperature entropy is negative in this case. This is definitely wrong
because the zero temperature entropy is by definition the logarithm of the
number of ground states, which in any case is at least $\ln 2$.  It turns out
that we have to be more carefull with the limit $T\to 0$ to get the correct zero
temperature entropy. In terms of $\Delta a$ the condition $\kappa > \kappa_c$
means
\begin{equation}
  \label{eq:what_it_means}
  2^{-N} > \Delta a \sqrt{\frac{2}{\pi N \ave{a^2}}}
\end{equation}
i.e.\ essentially $\Delta a = \bigo{2^{-N}}$. In this regime the contributions
of $\tilde S$ are $\bigo{N}$ for any finite $T$,
\begin{equation}
  \label{eq:contrib_tilde_S}
  \tilde S(\frac{\Delta a}T) = \ln(\frac{T}{\Delta a}) + 1 +
  \bigo{\frac{\Delta a^2}{T^2}},
\end{equation}
hence cannot be neglected.  Technically we deal with this contribution by
introducing an effective ``zero'' temperature $T_0$ below which the system can
not be ``cooled''. $T_0$ guarantees that the contribution of $\tilde S$ remains
$\bigo{N}$. Its value can be calculated from the lower bound of $S$:
\begin{displaymath}
  \ln 2 = N(\kappa_c-\kappa)\ln 2 + \tilde S(\frac{\Delta a}{T_0})
        \approx N(\kappa_c-\kappa)\ln 2 + \ln(\frac{T_0}{\Delta a}).
\end{displaymath}
From that we get
\begin{equation}
  \label{eq:T0}
  T_0 = 2\Delta a \, 2^{N(\kappa - \kappa_c)} = \sqrt{2\pi N \ave{a^2}}\,\,2^{-N}.
\end{equation}
In this regime the average ground state energy $\ave{E_1}$ is no longer 0 but
\begin{equation}
  \label{eq:E_kappa}
  \ave{E_1} = T_0 = \sqrt{2\pi N \ave{a^2}}\,\,2^{-N}
\end{equation}
This equation completes the previously known result that the average value of $E_1$ is
$\bigo{\sqrt{N}\,2^{-N}}$ for real valued input numbers
\cite{karmarkar:etal:86,lueker:98} by specifying the prefactor to be
$\sqrt{2\pi\ave{a^2}}$.

%% Figure here: E of N
\begin{figure}
  \includegraphics[width=\columnwidth]{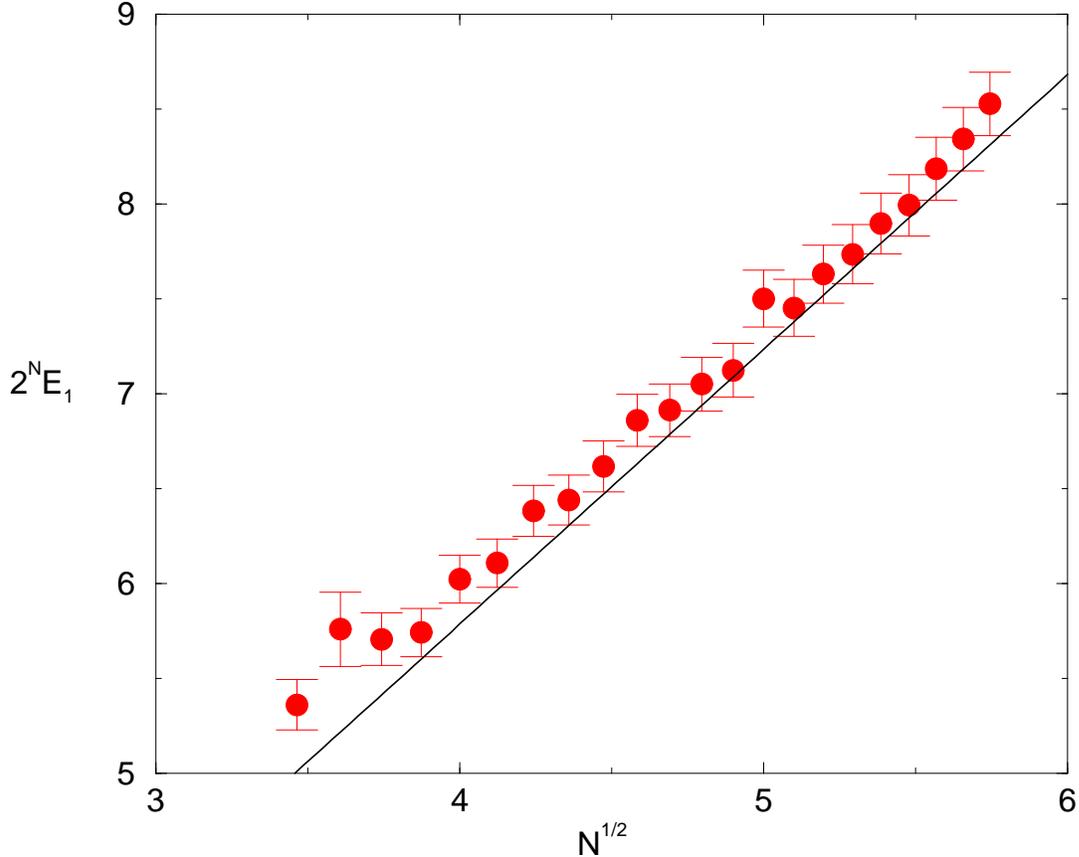}
\caption{\small
  Average solution of the number partitioning problem with input numbers $a_j$
  being \iid{} uniform on $[0,1]$ compared to the analytical result
  Eq.~\myref{eq:E_kappa} (straight line).  Each data point is the average over
  $10^4$ random samples.
\label{fig:eminuc_of_N}}
\end{figure}

To check Eq.~\myref{eq:E_kappa} we consider the continous variant of number
partitioning, where the $a_i$ are real numbers, uniformely distributed in the
interval $[0,1)$.  In our formalism this means $\Delta a \to 0$ and
$\sum\nolimits_ja_j^2=N/3$.  We are in the $\kappa > \kappa_c$ regime and
Eq.~\myref{eq:E_kappa} becomes
\begin{equation}
  \label{eq:E_kappa_cont}
  E_0 = \sqrt{\frac23 \pi N}\,\,2^{-N} = 1.447\,\sqrt{N}\,\,2^{-N}
\end{equation}
In Fig.~\myref{fig:eminuc_of_N}, Eq.~\myref{eq:E_kappa_cont} is compared to
numerical data. The agreement is convincing.

%% Figure here: Phase diagram
\begin{figure}
  \includegraphics[width=\columnwidth]{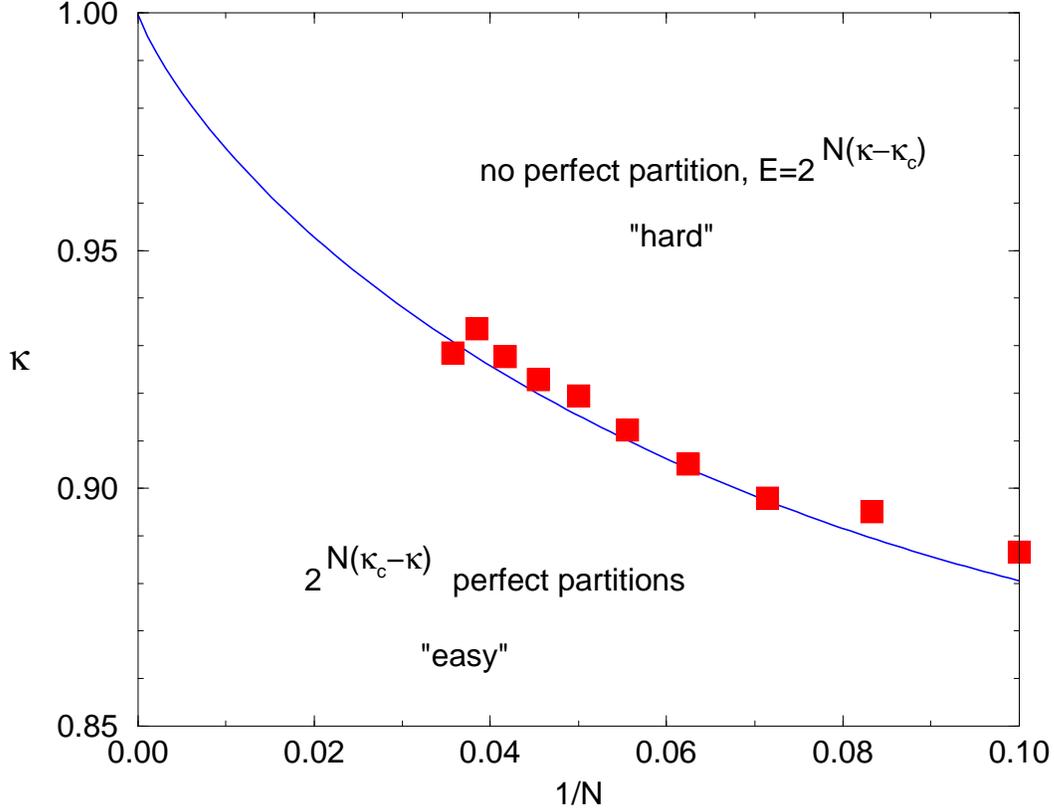}
\caption{\small
  Phase diagram of the random number partioning problem. $N\kappa$ is essentialy
  the number of significant bits to encode the input numbers, see
  Eq.~\myref{eq:kappa}. The squares denote the phase boundary found numerically.
  The solid line is given by $\kappa_c$ from Eq.~\myref{eq:kappa_c}.  For
  $\kappa<\kappa_c$, the zero temperature entropy is extensive and a search
  algorithm typically finds quickly one of the $O(2^N)$ perfect partitions. For
  $\kappa > \kappa_c$, no perfect partitions exist, and the optimization problem
  has a hard to find, unique solution.
\label{fig:kappa_of_N}}
\end{figure}

To check whether $\kappa(N)$ is a control parameter with a phase transition at
$\kappa_c(N)$, we did numerical simulations. For fixed $N$ and $\kappa$ we
calculated the fraction of instances that have at least one perfect partition.
In accordance with Gent and Walsh \cite{gent:walsh:96} we find that this
fraction is $1$ for small $\kappa$ and $0$ for larger $\kappa$. The transition
from $1$ to $0$ is sharp. Fig.~\myref{fig:kappa_of_N} shows the numerically
found transition points for $10 \leq N \leq 28$ compared to $\kappa_c(N)$ from
Eq.~\myref{eq:kappa_c}. Again the agreement is convincing. Note that
$\kappa_c(N\to\infty)=1$. The asymptotic estimate 0.96 given by Gent and Walsh
is probably due to the influence of the $\bigo{\frac1N\log N}$ term
in eq.~\myref{eq:kappa_c} which cannot be neglected for system sizes accesible for
simulations ($N \leq 30$).

Before we turn to the constrained NPP, let us summarize what we have found
so far: The statistical mechanics analysis of the NPP reveals two different
phases, distinguished by the value of a parameter $\kappa$, eq.~\myref{eq:kappa},
which corresponds to the number of significant bits in the encoding of the input
numbers $a_j$ divided by $N$.  For $\kappa < \kappa_c$, we have an exponential
number of perfect partition, hence an exponential number of solutions to the
NPP. For $\kappa > \kappa_c$, we only have 2 solutions with a partition
difference given by eq.~\myref{eq:E_kappa}.

\subsection{Statistical mechanics of the constrained NPP}
\label{sec:constrained-NPP}

The partition function of the constrained NPP is
\begin{equation}
  \label{eq:Z-constrained-def}
  Z = {\sum_{\{s_j\}}}' e^{-|\sum_ja_js_j|/T},
\end{equation}
where the primed sum denotes summation over all spin configurations with
$\sum_js_j=mN$. By now we know how to separate the absolute value on the
exponent from the summation variables, but here additionally we have to get rid
of the the constraint $\sum_js_j=mN$ to do the summation.  The discrete version
of the Dirac $\delta$ function, the Kronecker $\delta$ symbol
\begin{equation}
  \label{eq:def-kronecker}
  \delta_{n,m} = \int_{-\pi}^\pi \frac{d\hat{m}}{2\pi}e^{i\hat{m}(n-m)} =
  \left\{\begin{array}{ll}
    1 & \mbox{if $m = n$} \\
    0 & \mbox{if $m \neq n$}
  \end{array}\right.,
\end{equation}
for integer $m$, $n$ can be used to achieve this:
\begin{displaymath}
  Z = \sum_{\{s_j\}} \delta_{\sum_js_j,mN}e^{-\frac1T|\sum_ja_js_j|}
    = \int_{-\pi}^\pi \frac{d\hat{m}}{2\pi} e^{-i\hat{m}mN}
      \sum_{\{s_j\}} e^{i\hat{m}\sum_js_j -|\sum_ja_js_j|/T}
\end{displaymath}
The remaining sum can now be done exactly like in the preceeding section.  The
result is
\begin{equation}
  \label{eq:finally-Z}
  Z = 2^N \int_{-\pi/2}^{\pi/2}\frac{dy}{\pi}\int_{-\pi}^\pi\frac{d\hat{m}}{2\pi} 
  e^{N G(y,\hat{m})}
\end{equation}
with
\begin{eqnarray}
  \label{eq:def-G}
  G(y,\hat{m}) &=& i\hat{m}m + \frac{1}{N}\sum_{j=1}^N\ln\cos(\frac{a_j}{T}
  \tan y + \hat{m})\nonumber\\
  &\approx& i\hat{m}m + \ave{\ln{\cos(\frac{a}{T}\tan(y)+\hat{m})}}.
\end{eqnarray}
Compared to the unconstrained case we are left with a twofold integral and a
complex valued integrand. The generalization of the Laplace method to complex
integrands is the {\em saddle point method}
\cite{debruijn:61,mathews:walker:book}: Let the real part of $G(y,\hat{m})$ have
a maximum at $(y_0,\hat{m}_0)$. Then
\begin{eqnarray}
  \label{eq:saddle-point-example}
  \int e^{N G(y,\hat{m})} dy d\hat{m} &\approx& e^{N G(y_0,\hat{m}_0)} 
  \int\! dy d\hat{m} 
  e^{-\frac{N}{2}(y,\hat{m})\mathbf{G}(y,\hat{m})^T} \nonumber\\
  &=& e^{N G(y_0, \hat{m}_0)} \frac{2\pi}{N\sqrt{\det{\mathbf{G}}}}
\end{eqnarray}
for large $N$. $\mathbf{G}$ is the $2\times 2$ Hesse matrix
\begin{equation}
  \label{eq:def-hesse-matrix}
  \mathbf{G} = \left(
    \begin{array}{cc}
      \frac{\partial^2 G(y,\hat{m})}{\partial^2 y} & 
      \frac{\partial^2 G(y,\hat{m})}{\partial y \partial\hat{m}} \\
      \frac{\partial^2 G(y,\hat{m})}{\partial \hat{m} \partial y} & 
      \frac{\partial^2 G(y,\hat{m})}{\partial^2 y}
    \end{array}
  \right),
\end{equation}
where the derivatives are taken at the saddle point $(y_0,\hat{m}_0)$.  In our
case the saddle point equations are
\begin{equation}
  \label{eq:saddle-1-y}
  0 = \frac{\partial G(y,\hat{m})}{\partial y} = 
       \ave{\frac{a}{T}\tan(\frac{a}{T}\tan y + \hat{m})} (1+\tan^2 y)
\end{equation}
\begin{equation}
  \label{eq:saddle-1-m}
  0 = \frac{\partial G(y,\hat{m})}{\partial \hat{m}} = 
      \ave{\tan(\frac{a}{T}\tan y + \hat{m})} + i m.
\end{equation}
Like in the unconstrained case, we will assume that $a$ can only take on values
that are integer multiples of a fixed number $\Delta a$.  With this assumption
and the ansatz
\begin{equation}
  \label{eq:saddle-ansatz-y}
  \tan y_k = k\pi\frac{T}{\Delta a} + i T x \qquad k=0,\pm1,\pm2,\ldots
\end{equation}
\begin{equation}
  \label{eq:saddle-ansatz-m}
  \tilde{m} = -i\hat{m}
\end{equation}
the saddle point equations simplify to
\begin{equation}
  \label{eq:saddle-x}
  \ave{a\tanh(ax+\tilde{m})} = 0
\end{equation}
\begin{equation}
  \label{eq:saddle-m}
  \ave{\tanh(ax+\tilde{m})} = m.
\end{equation}
For given value of $m$, the saddle point equations yield a solution
$(x,\tilde{m})$, which in turn gives rise to an infinite number of saddle points
(eq.~\myref{eq:saddle-ansatz-y}). The contribution from all these saddle points have
to be summed up to give $Z$, the groundstate energy $E_1$ and the entropy $S$.
Unfortunately, for $m > 0$ we can solve the saddle point equations only
numerically.  Therefore we will for the time being concentrate on the {\em
  balanced} NPP, $m=0$.  In this case, the solution is trivial, $x=\tilde{m}=0$,
the determinant of the Hesse matrix is
\begin{equation}
  \label{eq:det-Hesse-balanced}
  \det\mathbf{G} = \frac{\ave{a^2}-\ave{a}^2}{T^2} \cdot 
   \left(1+\frac{\pi^2T^2}{\Delta a^2}k^2\right)^2
  \qquad k=0,\pm1,\pm2,\ldots.
\end{equation}
With the help of eq.~\myref{eq:coth-series-1} it is straightforward to sum up the
contributions from all saddle points. The result is:
\begin{equation}
  \label{eq:Z-balanced}
  Z = 2^N\cdot \frac{\Delta a}{N\pi\sqrt{\ave{a^2}-\ave{a}^2}} 
  \coth{\frac{\Delta a}{T}}
\end{equation}
The partition function for the balanced NPP is very similar to the one of the
unconstrained NPP, eq.~\myref{eq:Z-unconstrained}.  Only the denominator changes
from $\sqrt{\frac{\pi}{2}N\ave{a^2}}$ in the unconstrained to $\pi N
\sqrt{\ave{a^2}-\ave{a}^2}$ in the balanced case.  The discussion of entropy and
groundstate energy is very similar, too.  The entropy can be written as
\begin{equation}
  \label{eq:balanced-entropy}
  S = N (\kappa_c-\kappa) \ln 2 + \tilde{S}(\Delta a/T)
\end{equation}
where $\tilde{S}$ is the same as for the unconstrained NPP
(eq.~\myref{eq:thermal_entropy}) and the order parameter $\kappa$ and its critical
value $\kappa_c$ are
\begin{equation}
  \label{eq:balanced-kappa_c}
  \kappa_c = 1 - \frac{1}{N}\ln_2(\frac{\pi}{\sqrt{12}}N)
\end{equation}
\begin{equation}
  \label{eq:balanced-kappa}
  \kappa = \frac{1}{N}\ln_2(\frac{\sqrt{12}}{\Delta a}\sqrt{\ave{a^2}-\ave{a}^2}).
\end{equation}
The condition for the existence of perfect partitions, $\kappa < \kappa_c$,
translates into
\begin{equation}
  \label{eq:phase-boundary-special-balanced}
  b < N - \ln_2\big(\frac{\pi}{\sqrt{12}}N\big).
\end{equation}
for input numbers $a_j$ being \iid{} $b$-bit integers
(cf.~eq.~\myref{eq:phase-boundary-special}).  From fig.~\myref{fig:korf} one can
tell, that for $b=25$ $N$ must be larger than $30$ for perfect partitions to
exist. Eq.~\myref{eq:phase-boundary-special-balanced} yields $N > 29.75$.

%% Figure here: E of N balanced
\begin{figure}
  \includegraphics[width=\columnwidth]{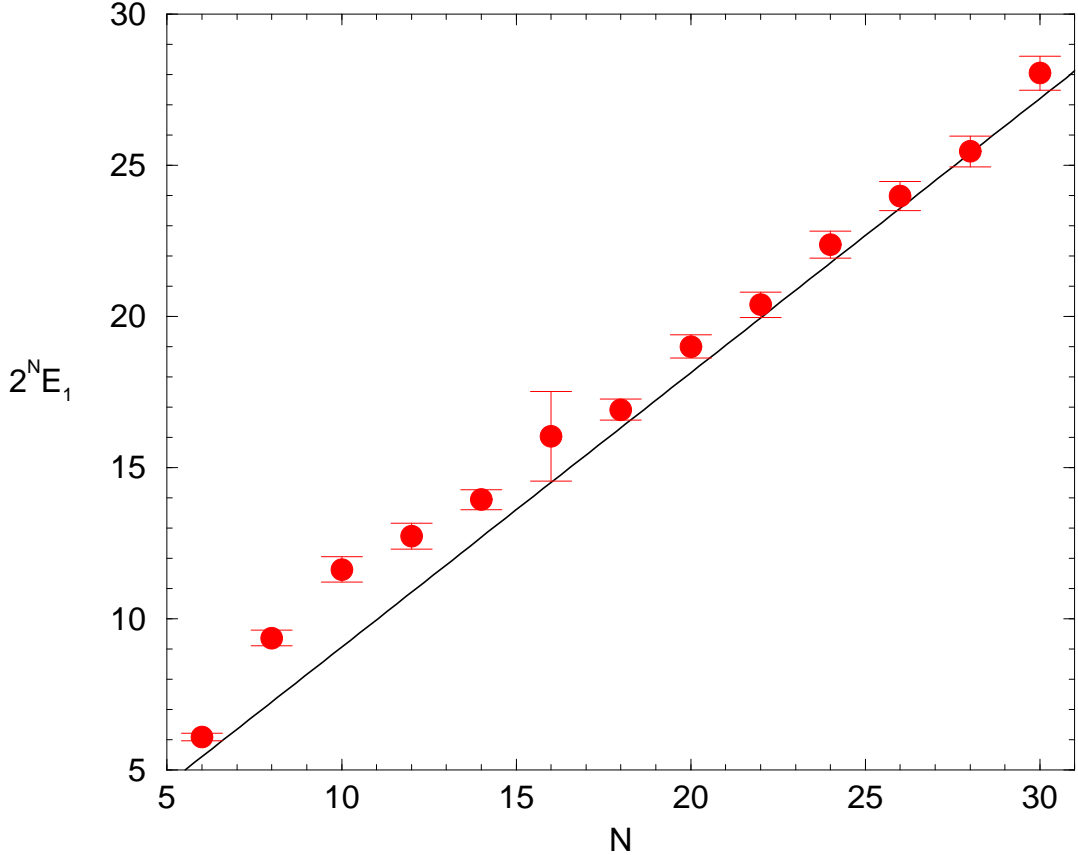}
\caption{\small
  Average minimum residue of the {\em balanced} number partitioning problem with
  real numbers $0\leq a_i < 1$ compared to the analytical result
  Eq.~\myref{eq:E_kappa-balanced} (straight line).  Each data point is the
  average over $10^4$ random samples.
\label{fig:emin_of_N}}
\end{figure}

For $\kappa > \kappa_c$, thee optimum partition difference $E_1$ reads
\begin{equation}
  \label{eq:E_kappa-balanced}
  \ave{E_1}= \pi\sqrt{\ave{a^2}-\ave{a}^2} \cdot N \cdot 2^{-N} 
\end{equation}
for the balanced NPP. Again this result fits very well with the numerics, see
fig.~\myref{fig:emin_of_N}.

\subsection{Overconstrained NPP}

\begin{figure}
  \includegraphics[width=0.45\columnwidth]{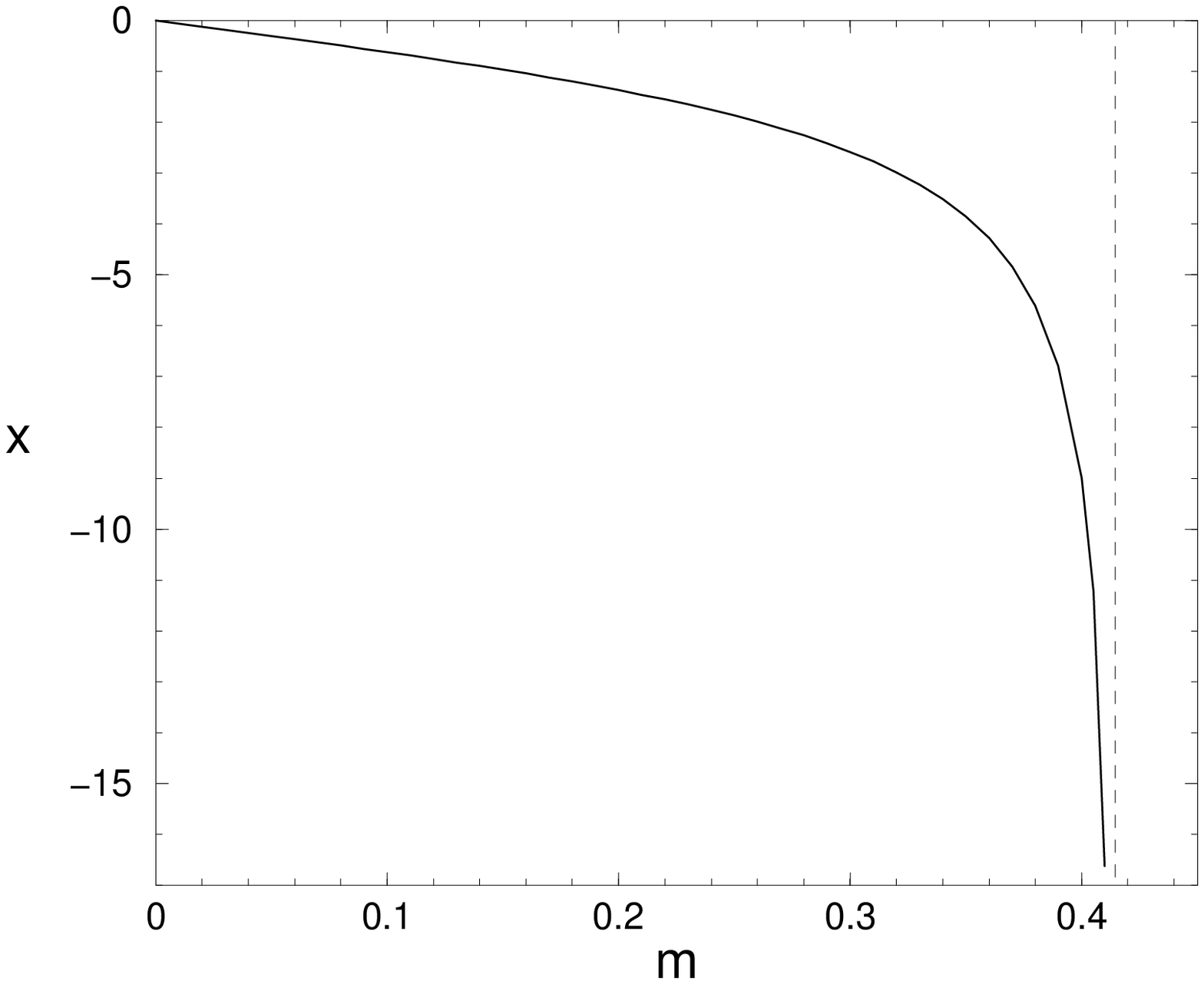} \hspace{0.05\columnwidth}
  \includegraphics[width=0.45\columnwidth]{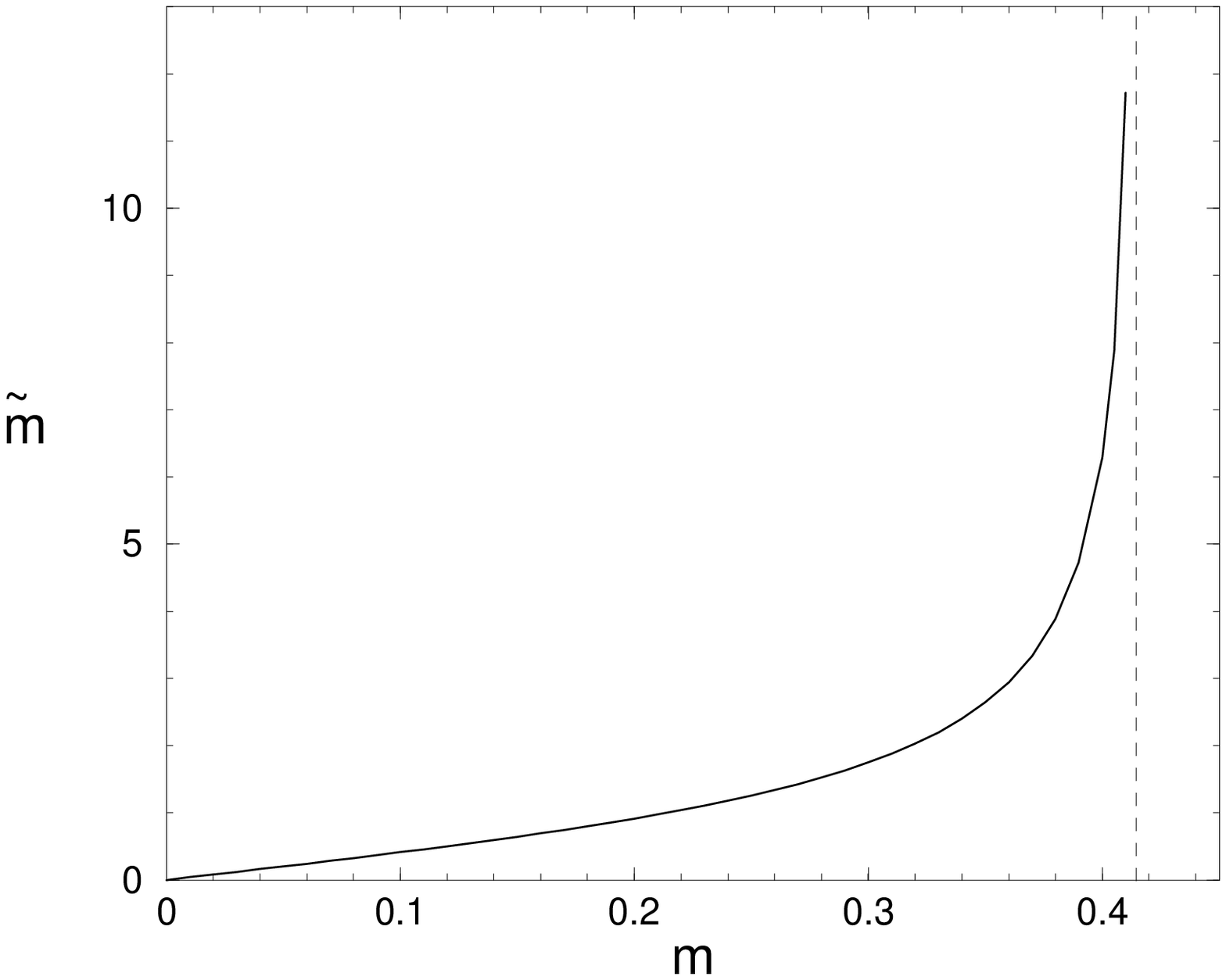}
\caption{\small
  Solution of the saddle point equations \myref{eq:saddle-x} and \myref{eq:saddle-m}
  for input numbers $a$ that are \iid{} uniform over $[0,1]$.
\label{fig:saddle-point-solution}}
\end{figure}

For $m>0$ we solve the saddle-point equations \myref{eq:saddle-x} and \myref{eq:saddle-m}
numerically.  Fig.~\myref{fig:saddle-point-solution} displays the
solution for input numbers $a$ that are \iid{} uniform over $[0,1]$.

%% Figure here: E of N balanced

The solution diverges if $m$ approaches a critical value $m_c = 0.41\ldots$. For
larger values of $m$, the saddle-point equations have no solution. This is no
surprise: $\tanh()$ is a monotonic function with $-1 \leq \tanh() \leq 1$.
Eq.~\myref{eq:saddle-x} requires that $\tanh(ax+\tilde{m})$ changes sign within
the integration range. Therefore the right hand side of eq.~\myref{eq:saddle-m}
has to be smaller than $1$. For input numbers distributed uniformely over
$[0,1]$ the saddle-point equations are
\begin{equation}
  \label{eq:saddle-x-uniform}
  \int_0^1 da \, a\tanh(ay+\tilde{m}) = 0
\end{equation}
\begin{equation}
  \label{eq:saddle-m-uniform}
  \int_0^1 da \,\tanh(ay+\tilde{m}) = m
\end{equation} 
and it is easy to show that $|m|$ has to be smaller than
$m_c=\sqrt{2}-1=0.41\ldots$ for a solution to exist.

In statistical mechanics, a diverging solution often indicates that the properties of the system
change drastically -- a phase transition. What kind of phase transition is
related to a critical value of the ``magnetization'' $m$? If $m$ is close to 1,
the NPP is {\em overconstrained}, i.e.\ we cannot expect to find a perfect
partition.  Instead, the optimum partition is the one that collects the
$\frac12(1-m)\cdot N$ {\em largest} numbers $a_j$ in the smaller subset. Let $a'$ be the
$\frac12(1+m)$ percentile of $a$,
\begin{equation}
  \label{eq:percentile}
  \frac12(1+m) = \int_0^{a'} \rho(a) da.
\end{equation}
The partition difference $E_s$ of this sorted partition then reads
\begin{equation}
  \label{eq:E-sorted}
  E_s/N = \int_0^{a'} a \rho(a) da - \int_{a'}^\infty a \rho(a) da.
\end{equation}
As long as $E_s > 0$ holds, the sorted partition is optimal. $E_s$ is positive
if $m$ is greater than a critical value $m_c$ that depends on the distribution
$\rho$.  For the uniform distribution on $[0,1]$, $a'=\frac12(1+m)$ and
\begin{equation}
  \label{eq:E-sorted-cont}
  E_s/N = \frac14(1+m)^2 - \frac12,
\end{equation}
which is positive provided
\begin{equation}
  \label{eq:m-crit-cont}
  m > m_c = \sqrt{2} - 1.
\end{equation}
A more complicated derivation of Eqs.~\myref{eq:E-sorted-cont} and
\myref{eq:m-crit-cont} using the statistical mechanics approach can be found in
\cite{ferreira:fontanari:98}.

Note that the computational complexity of the overconstrained NPP ($m > m_c$)
equals that of sorting $N$ numbers, i.e.\ $\bigo{N \ln N}$, so we have another
phase transition from a computationally hard to a computationally easy regime in
the NPP.

%%% Local Variables: 
%%% mode: latex
%%% TeX-master: "main"
%%% End: 
 \section{The random cost problem}
\label{sec:random}

The preceeding section has shown that the statistical mechanics of the NPP can
be analysed rather easily.  This is a remarkable exception. In general, spin
glass models are much harder to deal with, and physicists have considered
various simplifications.  One of these simplified models was Derrida's random
energy model, REM \cite{derrida:80,derrida:81}.  A cost function or Hamiltonian
like Eq.\myref{eq:spin-glass-hamiltonian} maps the random numbers
$J_{ij}$ onto $2^N$ random numbers $E_k$, distributed according to a probability
density $p(E)$.  Derrida's idea was to forget about the configurations $\{s_j\}$
and to consider directly the energies $E_k$ as {\em independent random numbers}, drawn
from the probablity density $p(E)$.  The essential simplification, which leads
to the analytic tractability of the model, is the assumption of statistical
independence.

The usefullness of the REM in spin glass theory has been discussed
elsewhere \cite{gross:mezard:84}. Here we will concentrate on its
counter part in combinatorial optimization, the {\em random cost problem}:
Given are $M$ random numbers
$E_k$, {\it independently} drawn from a density $p(E)$.
Find the minimum of these numbers.
Since every number has to be considered at least once, the computational
complexity of the random cost problem is $\bigo{M}$. The statistical independence
of the numbers prevents an efficient heuristic: Any heuristic algorithm 
that considers only $K\ll M$ numbers is no better than simple sequential 
search through an arbitrary $K$-element subset of the list.

The motivation to study random cost problems stems from the fact that every combinatorial
optimization problem with random inputs can be {\em approximated} by a random cost problem.
If the original optimization problem has $M$ feasible solutions and the costs of these
solutions are distributed with density $p(E)$, in the corresponding random cost problem we shall
simply assume, that the $M$ costs are drawn independently from $p(E)$. 
For the NPP the approximation by a random cost problem gives apparently correct
results at least for the statistics of the low cost configurations \cite{mertens:00a}.

\subsection{Distribution of costs}
\label{sec:distribution-energies}

To find a random cost problem that corresponds to the NPP, we first have
to calculate the probablity density of the costs. For the constrained NPP,
$p(E)$ reads
\begin{equation}
  \label{eq:def-rho}
  p(E) = {\binomial{N}{N_{+}}}^{-1}{\sum_{\{s_j\}}}'
  \ave{\delta(E-|\sum_ja_js_j|)}.
\end{equation}
where 
the primed sum runs over all configurations with 
$N_{+}=N\frac{1}{2}(1+m)$ spins $s_j = +1$.
Since the numbers $a_j$ are drawn independently from an identical
distribution, the average in Eq.~\myref{eq:def-rho} depends only on
$N_{+}$, but not on the particular spin configuration. Ignoring the
absolute value in the cost function for a moment, we may write
\begin{equation}
  \ave{\delta(E-\sum_ja_js_j)} = \int dy \, g_{N-N_{+}}(y) \, g_{N_{+}}(E+y),
\end{equation}
where $g_K$ is the probablity density of the sum $\sum_{j=1}^Ka_j$.
The central limit theorem tells us that for large $K$
\begin{equation}
  \label{eq:g-k}
  g_K(y) = \frac{1}{\sqrt{2\pi\sigma^2K}}
             \exp\left(-\frac{(y-K\ave{a})^2}{2\sigma^2K}\right),
\end{equation}
where $\sigma^2=\ave{a^2}-\ave{a}^2$ is the variance
of $a$. Hence
\begin{equation}
  \ave{\delta(E-\sum_ja_js_j)} = \frac{1}{\sqrt{2\pi\sigma^2N}}
             \exp\left(-\frac{(E-[2N_{+}-N]\ave{a})^2}{2\sigma^2N}\right).
\end{equation}
With $2N_{+}-N = mN$ and taking the absolute value of the cost function into
account we finally get
\begin{equation}
  \label{eq:p-of-E-constrained}
  p_{m}(E) = \frac{\Theta(E)}{\sqrt{2\pi\sigma^2N}}\left(
         e^{-\frac{(E-m\ave{a}N)^2}{2\sigma^2N}} + 
         e^{-\frac{(E+m\ave{a}N)^2}{2\sigma^2N}}
         \right)
\end{equation}
as the probablity density for the costs in the random, constrained
NPP. $\Theta(x)$ is the step function, $\Theta(x)=1$ for $x\geq0$,
$\Theta(x)=0$ for $x<0$.

To get the density of the costs for the unconstrained NPP, 
we have to sum over all values $N_{+}$,
\begin{displaymath}
  p(E) = 2^{-N} \sum_{N_{+}=0}^N \binomial{N}{N_{+}} p_{m=2N_{+}/N-1}(E).
\end{displaymath}
For large $N$, the sum is dominated by terms with $N_{+}=N\frac{1+m}{2}$,
$m=\bigo{1}$, and we may
apply the asymptotic expansion
\begin{equation}
  \label{eq:binomial-asymptotic}
  \binomial{N}{N\frac{1+m}{2}} \approx \frac{2^N}{\sqrt{\frac{1}{2}(1-m^2)\pi N}} e^{-N s(m)}
\end{equation}
with
\begin{equation}
  \label{eq:m-entropy}
  s(m) = \frac{1+m}{2}\ln(1+m) + \frac{1-m}{2}\ln(1-m),
\end{equation}
and we may replace the sum over $N_{+}$ by an integral over $m$,
\begin{displaymath}
  p(E) = \frac{\sqrt{N}}{\sqrt{2\pi}} \int_{-1}^{1} dm \, e^{-N s(m)} \, p_m(E).
\end{displaymath}
$s(m)$ has a maximum at $m=0$. Applying the Laplace method to evaluate
the $m$-integral for large $N$ we finally get
\begin{equation}
  \label{eq:p-of-E-unconstrained}
  p(E) = \frac{2\Theta(E)}{\sqrt{2\pi\ave{a^2}N}} e^{-\frac{E^2}{2\ave{a^2}N}}
\end{equation}
as the probability density in the random unconstrained NPP.

\subsection{Statistics of the optimum}
\label{sec:statistics-of-optimum}

We may now specify the random cost problem that corresponds to the
NPP: Given are $M$ random numbers $E_i$, {\em independently} drawn from the
density $p(E)$, eq.~\myref{eq:p-of-E-constrained} resp.\ 
eq.~\myref{eq:p-of-E-unconstrained}.  Find the minimum of these numbers.  To
connect to the NPP, $M$ is chosen to be
\begin{equation}
  \label{eq:choose-M}
  M=\frac12\binomial{N}{N(1+m)/2} \approx \frac{2^{N-1}}{\sqrt{\frac{\pi}{2}N(1-m^2)}}
  e^{-N s(m)}
\end{equation}
for the constrained NPP and $M=2^{N-1}$ for the unconstrained NPP.
Our claim is that the NPP is very well approximated by this random cost problem.

Let $E_k$ denote the $k$-th lowest cost of an instance of our random cost
problem. The probability density $\rho_1$ of the minimum $E_1$ can easily be
calculated:
\begin{equation}
  \label{eq:basic-rho1}
  \rho_1(E_1) = M\cdot P(E_1)\cdot \Big(1-\int_0^{E_1} P(E') dE'\Big)^{M-1}
\end{equation}
$E_1$ must be small to get a finite right-hand side in the large $M$ limit.
Hence we may write
\begin{eqnarray*}
  \rho_1(E_1) &\approx& M\cdot P(0) \cdot \Big(1-E_1 P(0)\Big)^{M-1} \\
              &\approx& M\cdot P(0) \cdot e^{-M P(0) E_1}.
\end{eqnarray*}
This means that the probability density of the scaled minimal cost,
\begin{equation}
  \label{eq:def-epsilon}
  \varepsilon_1 = M \cdot P(0) \cdot E_1
\end{equation}
for large $M$ converges to a simple exponential distribution,
\begin{equation}
  \label{eq:rho1}
  \rho_1(\varepsilon) = e^{-\varepsilon} \cdot \Theta(\varepsilon).
\end{equation}
Note that a rigorous derivation from eq.~\myref{eq:basic-rho1} to
eq.~\myref{eq:rho1} can be found in textbooks on extreme order statistics
\cite{galambos:book}.  Along similar lines one can show that the density
$\rho_k$ of the $k$-th lowest scaled cost is
\begin{equation}
  \label{eq:rhok}
  \rho_k(\varepsilon) = \frac{\varepsilon^{k-1}}{(k-1)!} \cdot e^{-\varepsilon} \cdot 
  \Theta(\varepsilon)
  \qquad k = 2,3,\ldots.
\end{equation}

Let us compare eqs.~\myref{eq:rho1} and \myref{eq:rhok} with known analytical and
numerical results for the NPP. From the moments of the exponential distribution
Eq.~\myref{eq:rho1}, $\ave{\varepsilon^n}=n!$, we get
\begin{equation}
  \label{eq:2nd-moment}
  r = \frac{\sqrt{\ave{E_1^2} - \ave{E_1}^2}}{\ave{E_1}} = 1
\end{equation}
for the relative width of the distribution,
in perfect agreement with the numerical findings for the NPP \cite{ferreira:fontanari:98}.
The average minimal cost is $\ave{E_1}=1/(M\cdot P(0))$, which gives
\begin{equation}
  \label{eq:avE1-constrained}
  \ave{E_1} = \pi\cdot\sigma\cdot N \cdot 2^{-N} \cdot e^{N(
             \frac{\ave{a}^2m^2}{2\sigma^2}+s(m))}
\end{equation}
for the constrained and
\begin{equation}
  \label{eq:avE1-unconstrained}
  \ave{E_1} =  \sqrt{2\pi\ave{a^2}}\cdot\sqrt{N}\cdot 2^{-N}
\end{equation}
for the unconstrained case. Again this is in
very good agreement with numerical simulations for the NPP \cite{ferreira:fontanari:98} 
and in perfect agreement with
our analytical results from the preceeding section, eqs.~\myref{eq:E_kappa} and
\myref{eq:E_kappa-balanced}. For the constrained case with $m > 0$, the minimal
cost increases with increasing $m$. This is reasonable, but nevertheless
eq.~\myref{eq:avE1-constrained} must be wrong for $|m| > 0$. For input numbers $a$
drawn uniformely from $[0,1]$ eq.~\myref{eq:avE1-constrained} predicts that
$\ave{E_1}$ is exponentially small as long as $|m| < 0.583\ldots$, but we know
from the preceeding section that for $m > m_c=\sqrt{2}-1=0.414\ldots$ the NPP is
overconstrained, hence has $\ave{E_1}=\bigo{N}$.

%% Figure here: rho1
\begin{figure}[thb]
  \includegraphics[width=\columnwidth]{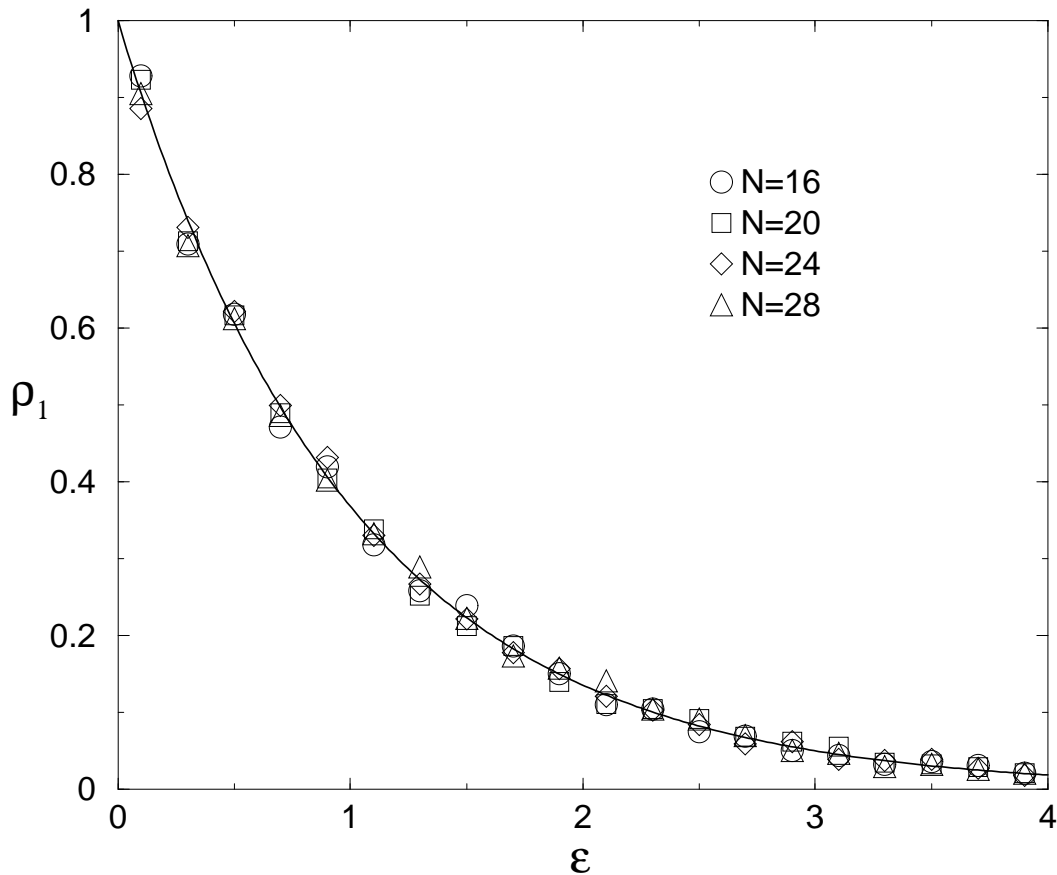}
\caption{\small 
  Distribution of the scaled optimum for the balanced number partioning problem.
  The solid line is given by Eq.~\myref{eq:rho1}, the symbols are averages over
  $10^4$ random samples.
  \label{fig:rho1}}
\end{figure}

%% Figure here: rhok
\begin{figure}[thb]
  \includegraphics[width=\columnwidth]{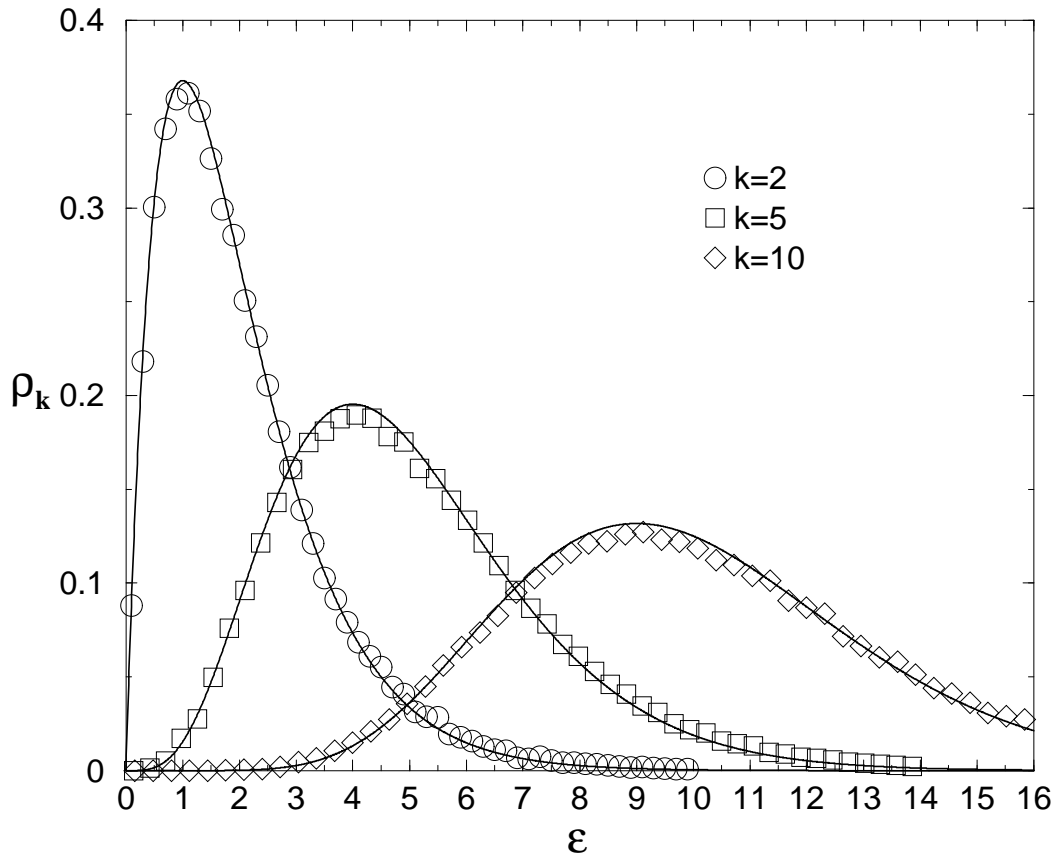}
\caption{\small 
  Distribution of scaled $k$-th lowest energy for the balanced number partioning
  problem. The solid lines are given by Eq.~\myref{eq:rhok}, the symbols are
  averages over $10^5$ random samples of size $N=24$.
\label{fig:rhok}}
\end{figure}

To check that the random cost approximation does not only give the correct first and
second moment of the optimum of the balanced NPP, we calculated the distribution
of $E_1$ and higher energies numerically. Figs.~\myref{fig:rho1} and
\myref{fig:rhok} display the results for the balanced NPP.  Equivalent plots for
the unconstrained NPP look similar.  The agreement between the numerical data
and Eqs.~\myref{eq:rho1} and \myref{eq:rhok} is convincing.

How can the random cost problem be so similar to the NPP? The answer is, that there is in
fact a certain degree of statistical independence among the costs in the NPP.
In the appendix we
show that the {\em joint probability} $p(E,E')$ factorizes, i.e.\ $p(E,E')=p(E)p(E')$
for the unconstrained and the balanced NPP, but not for the constrained NPP with $|m| > 0$.
This is a necessary, but not sufficient condition for independence, but the approach
can probably extended to a complete proof of independence. Here we adopt a physicists attitude
and consider the random cost problem to be a very good approximation to the NPP.

\subsection{Poor performance of heuristic algorithms}
\label{sec:bad-performance}

The correspondence between the NPP and the random cost problem not only provides 
analytic results on the NPP but also has some consequences for the dynamics of
algorithms: Any heuristic that exploits a fraction of the domain, generating and
evaluating a series of feasible configurations, cannot be significantly
better than random search.  
The best solution found by random search is distributed according to
Eq.~(\myref{eq:basic-rho1}), i.e.\ the average heuristic solution should approach
the true optimum no faster than $\bigo{1/M}$, $M$ being the number of configurations
generated.  Note that the best known heuristic,
Korf's CKK \cite{korf:98,mertens:99a} converges slower,
namely, like $\bigo{1/M^\alpha}$ with $\alpha < 1$ to the true optimum. 
Other heuristics, like simulated annealing, are even worse \cite{johnson:etal:91}.

%% Figure here: kk-discrepance
\begin{figure}[thb]
  \includegraphics[width=\columnwidth]{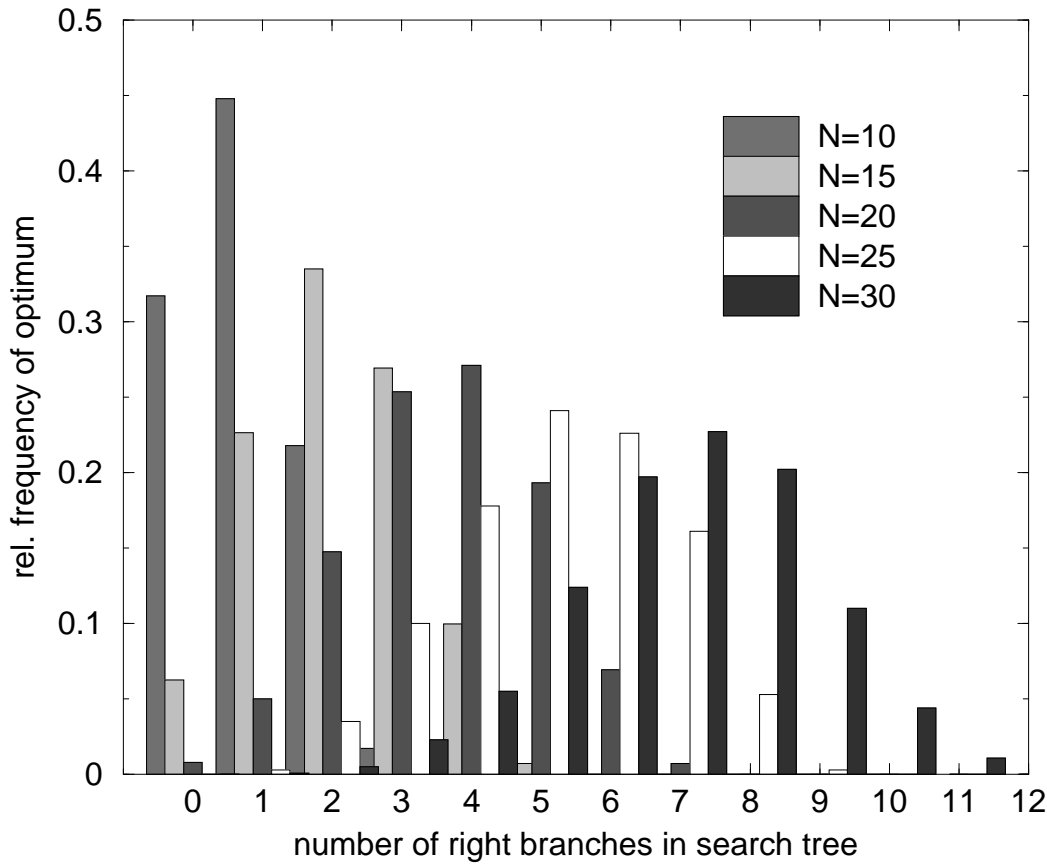}
\caption{\small \label{fig:kk-discrepance}
  Distribution of optima in the search tree spanned by the CKK algorithm. The number
  of right branches, the {\em discrepancy}, is a measure of distance
  to the heuristic Karmarkar-Karp solution. The NPP solved was unconstrained, with
  input numbers drawn uniformely from $[0,1]$. The number of instances is $10^4$ for
  $N\leq 20$ and $10^3$ otherwise.
}
\end{figure}

The random cost analogy means that there is hardly any correlation between a partition
and its cost. Partitions that are similar to each other may have very different costs and
vice versa.
One might argue that this picture depends on the {\em encoding} of a partition,
especially on the precise definition of ``similarity''. Throughout this paper we
used the obvious encoding of a partition as a set of binary variables and one can show
that indeed partitions with similar costs are completely dissimilar in terms
of a vanishing overlap, $\frac{1}{N}\sum_js_js'_j = 0$ (see appendix).

Maybe there is a better problem representation for the NPP, an encoding
that centers the good solution around a known position in search space.
In fact it has been found that the choice of encoding is more important than the choice
of search technique in determining search efficacy \cite{ruml:etal:96}. None of the proposed
encodings and search techniques is more efficient than Korf's CKK, however.

%% Figure here: kk-dist average and variance
\begin{figure}
  \includegraphics[width=0.45\columnwidth]{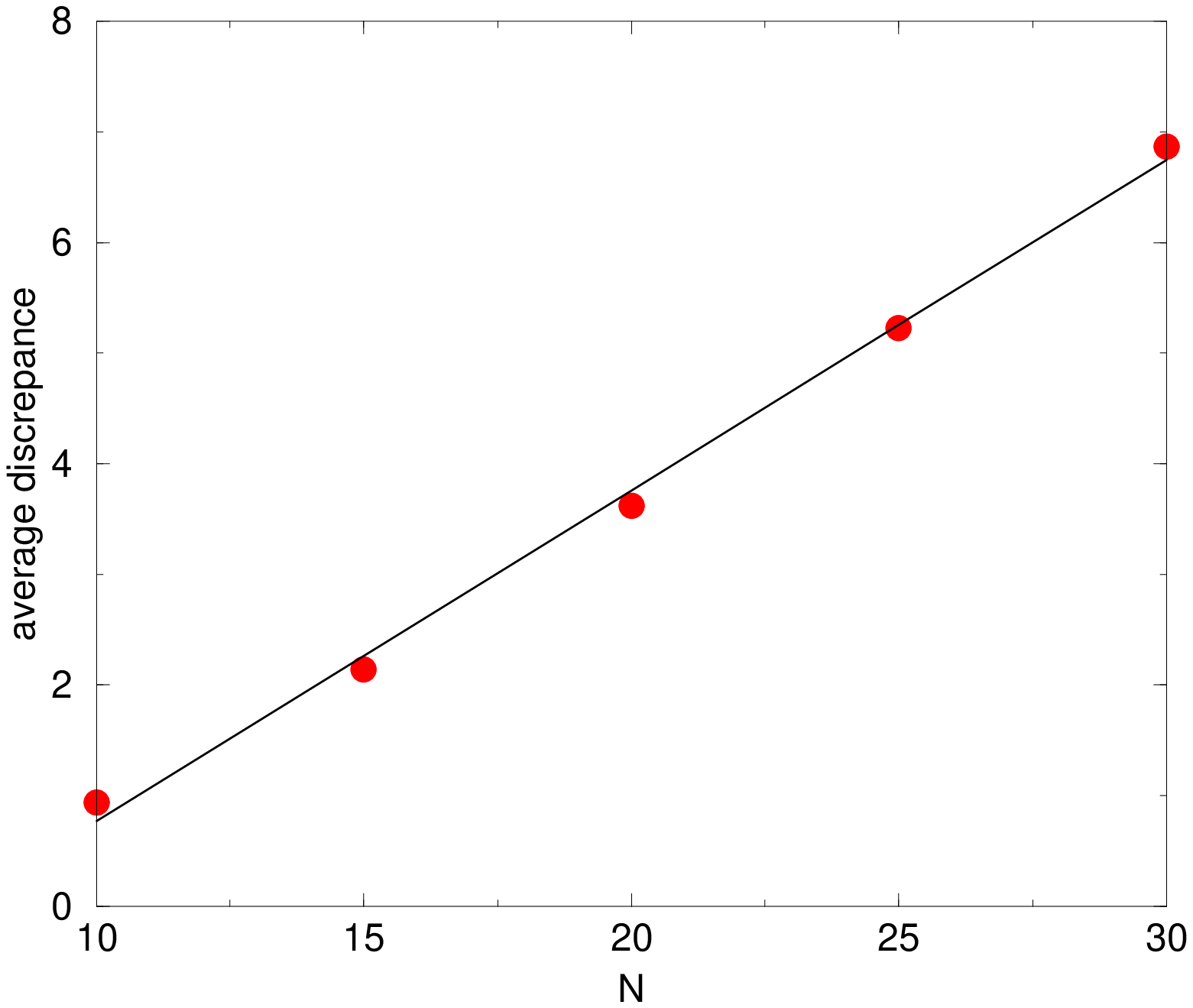} \hspace{0.05\columnwidth}
  \includegraphics[width=0.45\columnwidth]{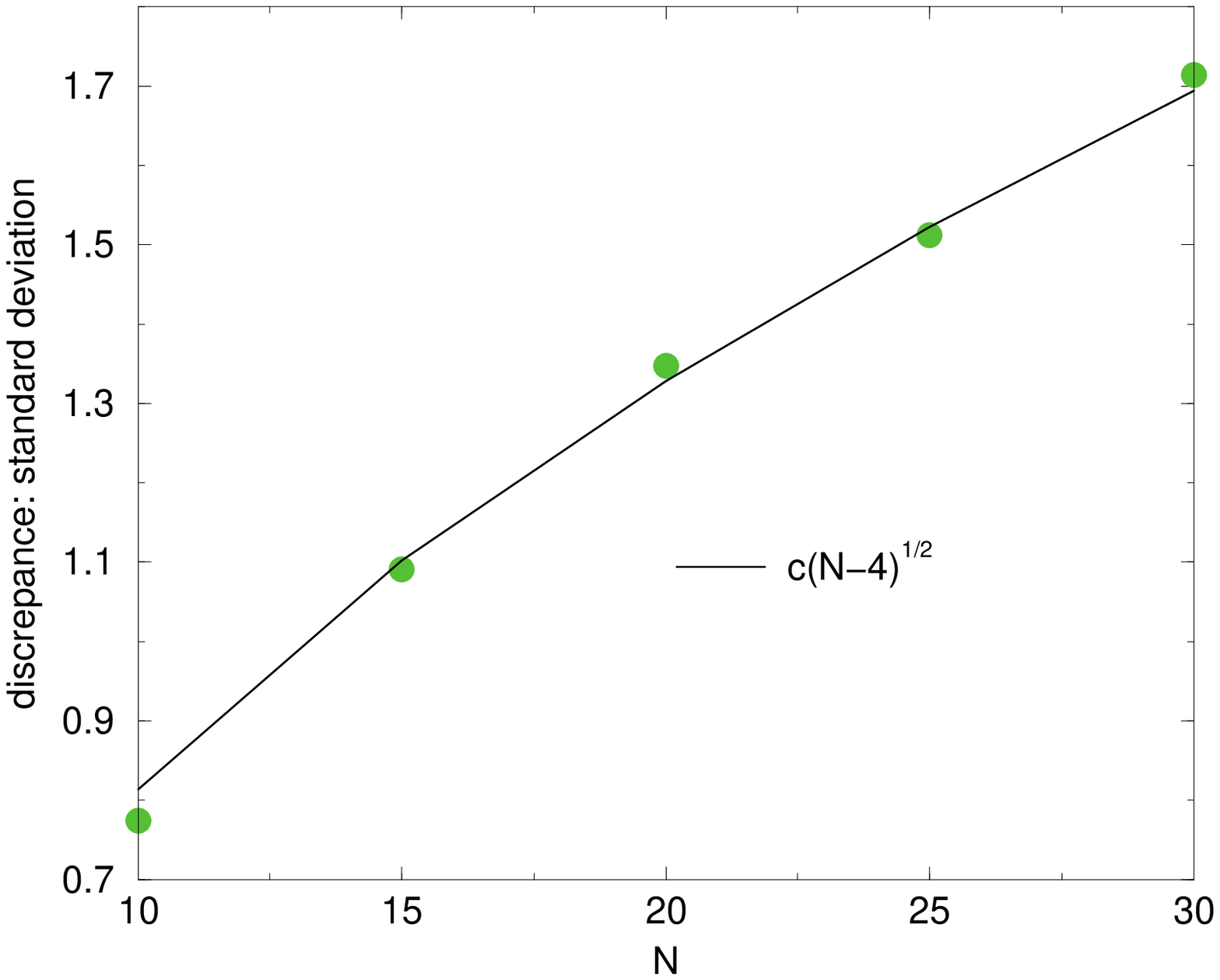}
\caption{\small \label{fig:kkdist-av-si}
  Discrepancy of the optimum partition: Mean (left) and standard deviation (right)
  vs.~$N$ from the data of fig.~\myref{fig:kk-discrepance}. The fit of $\sqrt{N-4}$ 
  for the standard deviation takes into account that the actual search tree
  is only explored to depth $N-4$, since for $N \leq 4$ the Karmarkar-Karp solution is 
  always optimal \cite{gent:walsh:98}. 
}
\end{figure}

An ansatz to concentrate good solutions in a small part in search space was proposed
by Korf \cite{korf:98}.
In the tree spanned by the CKK, a left branch follows the Karmarkar-Karp differencing heuristic,
a right branch does the opposite. Instead of searching the tree
depth first like in the CKK one might as well search the paths of the tree in increasing order
of the number of right branches, or discrepancies, from the heuristic recommendations. The hope
behind this {\em limited discrepancy search} 
\cite{harvey:ginsberg:95} is of course that good solutions
are ``close'' to the Karmarkar-Karp heuristic solution.

To check whether the discrepancy is a good measure to search the optimum, we calculated
the discrepancy of the optimum partition numerically. Figs.~\myref{fig:kk-discrepance}
and \myref{fig:kkdist-av-si} show that for large $N$, the discrepance of the optimum 
is a Gaussian distributed random variable with 
mean $\propto N$ and variance $\propto \sqrt{N}$. Hence a search guided by small discrepancies is
not significantly better than simple random search and our random cost analogy persists in  
discrepancy space. Note however that there are {\em some} correlations
between the optimum and its discrepancy. For true random discrepancies we would get an average of
$\frac{1}{2}N$, but from fig.~\myref{fig:kkdist-av-si} we get a smaller value ofabout $\frac{3}{10}N$.
This is no surprise: the partition with the highest discrepance $N-1$ has maximum cost, and
partitions with high discrepancy are never optimal. Hence the relevant range of discrepancies is
not $N$ but a fraction $xN$, $x < 1$. The best a heuristic search can do is to avoid the very
large discrepancies. Note that our independent cost assumption does not hold for costs
$\bigo{N}$ (see appendix).

%%% Local Variables: 
%%% mode: latex
%%% TeX-master: "main"
%%% End: 
 \section{Summary and conclusions}
\label{sec:summary}

The main contribution of this paper is the application of methods and ideas from
statistical mechanics to the number partitioning problem. 
The typical computational complexity of the NPP undergoes a sudden change if the
system size and/or the number of bits
in the input numbers is varied across a critical value. The standard statistical mechanics
approach yields a quantitative, analytic theory of this phasetransition.
Furthermore, it reveals a second phasetransition in the constrained NPP, controled by the 
imposed cardinality difference $m$.

The idea of a random energy model, born in spin glass theory, can be reformulated
as random cost problem in combinatorial optimization. Numerical as well as analytical
results support our claim, that the balanced and the unconstrained
NPP are extremely well approximated by a random cost problem, at least if one excludes 
the very high
costs. On the one hand this correspondence is responsible for the bad performance of
heuristic search algorithms but on the other hand allows
to derive new anayltic results like the probability density of optimal and suboptimal
costs.

What contributions does this work make beyond the specific problem of number partitioning?
First, it provides an example of the applicability of statistical mechanics to combinatorial
optimization that is exceptionally simple. It may well serve as a pedagogical introduction
into this interdisciplinary field. Second the idea of the random cost problem may have
applications beyond number partitioning.  The dynamics of heuristic
algorithms for other combinatorial optimization problems can be checked for a
signature of a corresponding random cost problem, possibly with a differing $p(E)$.

%%% Local Variables: 
%%% mode: latex
%%% TeX-master: "main"
%%% End: 

\appendix
\section{Evidence for the independent cost assumption}
\label{sec:independence}

Let $p(E,E')$ denote the joint probability density of having costs $E$ and $E'$ in the
same instance of an NPP,
\begin{equation}
  \label{eq:def-p-of-E-E'}
  p(E,E') = {\binomial{N}{N_{+}}}^{-2}{\sum_{\{s_j\}}}'{\sum_{\{s'_j\}}}'
  \ave{\delta(E-|\sum_ja_js_j|) \cdot \delta(E'-|\sum_ja_js'_j|)}.
\end{equation}
If the costs were independent, this
probabilty density should factorize, $p(E,E')=p(E)\cdot p(E')$.  This is what we
are going to check in this section. Our line of reasoning is similar to the proof
of theorem 4.9 in \cite{coffman:lueker:91}.

Consider the quantity
\begin{equation}
  \label{eq:def-p-tilde}
  \tilde{p}_{N_{++}}(E, E') :=
   \ave{\delta(E-\sum_ja_js_j) \cdot
    \delta(E'-\sum_ja_js'_j)}.
\end{equation}
All indices $j$ are treated equally in the average over the $a_j$,
hence $\tilde{p}$ can depend on $\{s_j\}$ and $\{s'_j\}$ only through the 
number $N_{++}$ of spins $s_j=s'_j=+1$. Hence
we can write
\begin{multline}
  \label{eq:p-of-E-E'-q}
  p(E,E') = \Theta(E)\Theta(E'){\binomial{N}{N_{+}}}^{-1}\sum_{N_{++}=0}^{N_{+}} 
  \binomial{N_{+}}{N_{++}} \binomial{N-N_{+}}{N_{+}-N_{++}} \\
  \cdot\bigg(\tilde{p}_{N_{++}}(E,E')+\tilde{p}_{N_{++}}(-E,E')+
        \tilde{p}_{N_{++}}(E,-E')+\tilde{p}_{N_{++}}(-E,-E')\bigg)
\end{multline}
The four $\tilde{p}$ terms and the $\Theta$-functions
take into account the absolute value of the cost-function
which we have omitted in the definition of $\tilde{p}$.
There are $\binomial{N}{N_{+}}$ possible ways to
choose the $+1$ spins in $\{s_j\}$. This factor cancels one of the
normalization factors in eq.~\myref{eq:def-p-of-E-E'}. 
Among the $N_{+}$ $+1$ spins in $\{s_j\}$ we can
choose $N_{++}$ spins that are $+1$ in $\{s'_j\}$, too. The remaining
$N_{+}-N_{++}$ $+1$ spins in $\{s'_j\}$ can be chosen among the
$N-N_{+}$ spins that are $-1$ in $\{s_j\}$. This yields the two binomial factors
in eq.~\myref{eq:p-of-E-E'-q}. 
Let
\begin{displaymath}
  {\mathcal A} = \{j : s_j=+1\} \qquad \overline{\mathcal A} = \{j : s_j=-1\}
\end{displaymath}
\begin{displaymath}
  {\mathcal A'} = \{j : s'_j=+1\} \qquad \overline{\mathcal A'} = \{j : s'_j=-1\}
\end{displaymath}
be the partitions corresponding to both spin sequences. 
With
\begin{equation}
  \label{eq:def-V1}
  V_1 := \sum_{j\in{\mathcal A}\cap{\mathcal A'}} a_j - 
        \sum_{j\in\overline{\mathcal A}\cap\overline{\mathcal A'}} a_j
\end{equation}
\begin{equation}
  \label{eq:def-V2}
  V_2 := \sum_{j\in{\mathcal A}\cap\overline{\mathcal A'}} a_j - 
        \sum_{j\in\overline{\mathcal A}\cap{\mathcal A'}} a_j
\end{equation}
we can write
\begin{equation}
  \label{eq:a-by-V}
  \sum_{j=1}^N a_js_j = V_1 + V_2 \qquad \sum_{j=1}^N a_js'_j = V_1 - V_2.
\end{equation}
Now the nice thing about $V_1$ and $V_2$ is that they depend on two disjoint
subsets of the numbers $a_j$, hence are statistically independent.  Let $\rho_1$
and $\rho_2$ denote the probabilty density of $V_1$ resp.\ $V_2$.  We may write
\begin{equation}
  \label{eq:delta-delta-rho-rho}
  \tilde{p}_{N_{++}} =
     \frac12 \cdot \rho_1\left(\frac12[E+E']\right)\cdot
      \rho_2\left(\frac12[E-E']\right).
\end{equation}
$V_1$ and $V_2$ are composed of sums of independent random numbers.  The first
sum in $V_1$ runs over all elements that have $s_j=s'_j=+1$, the second over
those with $s_j=s'_j=-1$.  Counting the number of elements in these sets,
\begin{eqnarray*}
  |{\mathcal A}\cap{\mathcal A'}| &=& N_{++} \\
  |{\mathcal A}\cap\overline{\mathcal A'}| &=& N_{+} - N_{++} =: N_{+-}\\
  |\overline{\mathcal A}\cap{\mathcal A'}| &=& N_{+} - N_{++} =: N_{-+}\\
  |\overline{\mathcal A}\cap\overline{\mathcal A'}| &=& N - 2N_{+}+N_{++} =: N_{--},
\end{eqnarray*}
we can write
\begin{eqnarray}
  \label{eq:rho-g}
  \rho_1(V_1) &=& \int dz \, g_{N_{++}}(z+V_1) \, g_{N_{--}}(z) \\
  \rho_2(V_2) &=& \int dz \, g_{N_{+-}}(z+V_2) \, g_{N_{-+}}(z),
\end{eqnarray}
where $g_k$ is the probabilty density of the sum of $k$ numbers $a_j$.  As
argued above, $\tilde{p}_{N_{++}}(E,E')$
depends on the spin sequences only through $N_{++}$,
\begin{eqnarray}
  \label{eq:delta-delta-N++}
   \tilde{p}_{N_{++}}(E,E')&=&
     \frac{1}{2} \int dz \, g_{N_{++}}\left(z+\frac12[E+E']\right) \, g_{N_{--}}(z)  
      \nonumber\\
             & & \cdot \int dz \, g_{N_{+-}}\left(z+\frac{1}{2}[E-E']\right) \, 
                  g_{N_{-+}}(z)
\end{eqnarray}

Now we consider the large $N$ limit. We are interested in the case $m=\bigo{1}$, hence
$N_{++}=\bigo{N}$ and 
the sum over $N_{++}$ in eq.~\myref{eq:p-of-E-E'-q} is dominated by contributions
with $N_{++}=\bigo{N}$. 
It is convenient to express $N_{++}$, $N_{--}$, $N_{-+}$ and
$N_{+-}$ in terms of the {\em overlap} parameter
\begin{equation}
  \label{eq:def-q}
  q := \frac{1}{N}\sum_{j=1}^N s_js'_j,
\end{equation}
which is $\bigo{1}$ in this scaling regime:
\begin{eqnarray*}
  N_{++} &=& \frac{N}{2}(1+m)-\frac{N}{4}(1-q) \\
  N_{--} &=& \frac{N}{2}(1+m)-\frac{N}{4}(1-q) \\
  N_{-+} &=& N_{+-} = \frac{N}{4}(1-q).
\end{eqnarray*}
Approximating all distributions $g$ in
eq.~\myref{eq:delta-delta-N++} by their asymptotic 
expansions, eq.~\myref{eq:g-k}, we get
\begin{multline}
  \label{eq:tilde-p-q}
  \tilde{p}_{N_{++}}(E,E') = \frac{1}{2\pi\sigma^2N\sqrt{1-q^2}}
   \exp\left(-\frac{(E-\ave{a}mN)^2}{2\sigma^2N(1-q^2)}
             -\frac{(E'-\ave{a}mN)^2}{2\sigma^2N(1-q^2)}\right) \cdot\\
   \exp\left(-q\frac{(E+E')\ave{a}mN - E E' - \ave{a}^2m^2N^2}{\sigma^2N(1-q^2)}\right)
\end{multline}
Note that $\tilde{p}_{N_{++}}(E,E')$ factorizes only for $q=0$.

In the scaling regime $m=\bigo{1}$, $q=\bigo{1}$ and $N$ large, we
can replace the binomials in eq.~\myref{eq:p-of-E-E'-q} by their asymptotic
expansions according to eq.~\myref{eq:binomial-asymptotic} and the sum over 
$N_{++}$ by an integral over $q$. This integral in turn can be calculated asymptotically
using the Laplace method. As a matter of fact the product of the binomials,
\begin{equation}
  \label{eq:binomial-product}
   \binomial{N_{+}}{N_{++}}\cdot\binomial{N-N_{+}}{N_{+}-N_{++}} =
   \binomial{\frac{N}{2}(1-m)}{\frac{N}{4}(1-q)}\cdot
   \binomial{\frac{N}{2}(1+m)}{\frac{N}{4}(1-q)}
\end{equation}
has a maximum at $q=m^2$, hence we expect the factorization only for $m=0$, the balanced NPP.
In fact, in this case we get
\begin{equation}
  \label{eq:sum-to-integral}
  \binomial{N}{N/2} \sum_{N_{++}=0}^{N/2}{\binomial{N/2}{N/2-N_{++}}}^2 \cdots =
  \frac{\sqrt{N}}{\sqrt{2\pi}}\int\frac{dq}{1-q^2}e^{-Ns(q)} \cdots
\end{equation}
with $s(q)$ from eq.~\myref{eq:m-entropy}, which has a maximum at $q=0$. This proves
that $p(E,E')$ factorizes asymptotically for the balanced NPP. Note however, that
even for $m=0$ the saddlepoint is not at $q=0$ if $E,E'=\bigo{N}$. Costs this large
are not independent.

For the unconstrained NPP the calculation of $p(E,E')$ is a bit more cumbersome, but similar.
Here we have two additional integrals over $m=\frac{1}{N}\sum_js_j$ and $m'=\frac{1}{N}\sum_js'_J$,
but in the limit $N\to\infty$ the major contributions come from the saddle point at $m=m'=0$.
Hence we observe the asymptotic factorization for the unconstrained NPP, too.

\bibliographystyle{plain}

\bibliography{complexity,cs,physics,math}

\begin{thebibliography}{10}

\bibitem{barahona:82}
F.~Barahona.
\newblock On the computational complexity of {I}sing spin glass models.
\newblock {\em J.\ Phys.\ A}, 15:3241, 1982.

\bibitem{debruijn:61}
N.G.~De Bruijn.
\newblock {\em Asymptotic Methods in Analysis}.
\newblock {John Wiley \& Sons}, {New York}, 1961.

\bibitem{cheeseman:etal:91}
Peter Cheeseman, Bob Kanefsky, and William~M. Taylor.
\newblock Where the {\em really} hard problems are.
\newblock In J.~Mylopoulos and R.~Rediter, editors, {\em Proc. of IJCAI-91},
  pages 331--337, San Mateo, CA, 1991. Morgan Kaufmann.

\bibitem{coffman:lueker:91}
E.G. Coffman and George~S. Lueker.
\newblock {\em Probabilistic Analysis of Packing and Partitioning Algorithms}.
\newblock John Wiley \& Sons, New York, 1991.

\bibitem{derrida:80}
Bernard Derrida.
\newblock Random-energy model: Limit of a family of disordered models.
\newblock {\em Phys.\ Rev.\ Lett.}, 45:79--82, 1980.

\bibitem{derrida:81}
Bernard Derrida.
\newblock Random-energy model: An exactly solvable model of disordered systems.
\newblock {\em Phys.\ Rev.\ E}, 24(5):2613--2626, 1981.

\bibitem{ferreira:fontanari:98}
F.F. Ferreira and J.F. Fontanari.
\newblock Probabilistic analysis of the number partitioning problem.
\newblock {\em J.\ Phys.\ A}, 31:3417--3428, 1998.

\bibitem{fu:89}
Yaotian Fu.
\newblock The use and abuse of statistical mechanics in computational
  complexity.
\newblock In Daniel~L. Stein, editor, {\em Lectures in the Sciences of
  Complexity}, volume~1, pages 815--826, Reading, Massachusetts, 1989.
  Ad\-di\-son-Wes\-ley Pub\-lish\-ing Company.

\bibitem{galambos:book}
J\'anos Galambos.
\newblock {\em The Asymptotic Theory of Extreme Order Statistics}.
\newblock Robert E.~Krieger Publishing Co., Malabar, Florida, 1987.

\bibitem{garey:johnson:79}
Michael~R. Garey and David~S. Johnson.
\newblock {\em Computers and Intractability. A Guide to the Theory of
  {NP}-Completeness}.
\newblock W.H.~Freeman, New York, 1997.

\bibitem{gent:walsh:95}
Ian~P. Gent and Toby Walsh.
\newblock Phase transitions from real computational problems.
\newblock In {\em Proceedings of the 8th International Symposium on Artificial
  Intelligence}, pages 356--364, Monterrey, 1995. ITESM.

\bibitem{gent:walsh:96}
Ian~P. Gent and Toby Walsh.
\newblock Phase transitions and annealed theories: Number partitioning as a
  case study.
\newblock In W.~Wahlster, editor, {\em Proc.~of ECAI-96}, pages 170--174, New
  York, 1996. John Wiley \& Sons.

\bibitem{gent:walsh:98}
Ian~P.\ Gent and Toby Walsh.
\newblock Analysis of heuristics for number partitioning.
\newblock {\em Computational Intelligence}, 14(3):430--451, 1998.

\bibitem{gross:mezard:84}
D.J.\ Gross and M.\ M\'ezard.
\newblock The simplest spin glass.
\newblock {\em Nucl.\ Phys.\ B}, 240:431--452, 1984.

\bibitem{harvey:ginsberg:95}
W.D. Harvey and M.L. Ginsberg.
\newblock Limited discrepancy search.
\newblock In {\em Proceedings IJCAI-95}, pages 607--613, Montreal, Quebec,
  1995.

\bibitem{johnson:etal:91}
David~S. Johnson, Cecicilia~R. Aragon, Lyle~A. McGeoch, and Catherine Schevron.
\newblock Optimization by simulated annealing: an experimental evaluation; part
  {II}, graph coloring and number partitioning.
\newblock {\em Operations Research}, 39(2):378--406, May-June 1991.

\bibitem{karmarkar:karp:82}
Narendra Karmarkar and Richard~M. Karp.
\newblock The differencing method of set partitioning.
\newblock Technical Report UCB/CSD 81/113, Computer Science Division,
  University of California, Berkeley, 1982.

\bibitem{karmarkar:etal:86}
Narendra Karmarkar, Richard~M. Karp, George~S. Lueker, and Andrew~M. Odlyzko.
\newblock Probabilistic analysis of optimum partitioning.
\newblock {\em J. Appl. Prob.}, 23:626--645, 1986.

\bibitem{kirkpatrick:gelatt:vecchi:83}
S.~Kirkpatrick, Jr. C.D.~Gelatt, and M.P. Vecchi.
\newblock Optimization by simulated annealing.
\newblock {\em Science}, 220(4598):671--680, May 1983.

\bibitem{korf:95}
Richard~E. Korf.
\newblock From approximate to optimal solutions: A case study of number
  partitioning.
\newblock In Chris~S. Mellish, editor, {\em Proc. of IJCAI-95}, pages 266--272,
  San Mateo, CA, 1995. Morgan Kaufmann.

\bibitem{korf:98}
Richard~E. Korf.
\newblock A complete anytime algorithm for number partitioning.
\newblock {\em Artificial Intelligence}, 106:181--203, 1998.

\bibitem{lighthill:59}
M.J.\ Lighthill.
\newblock {\em Introduction to Fourier Analysis and Generalised Functions}.
\newblock {Cambridge University Press}, {}, 1959.

\bibitem{lueker:98}
George~S.\ Lueker.
\newblock Exponentially small bounds on the expected optimum of the partition
  and subset sum problems.
\newblock {\em Random Structures and Algorithms}, 12:51--62, 1998.

\bibitem{mathews:walker:book}
Jon Mathews and R.L.\ Walker.
\newblock {\em Mathematical Methods of Physics}.
\newblock Addison-Wesley Publishing Company, Inc., Redwood City, California,
  1970.

\bibitem{mertens:98a}
Stephan Mertens.
\newblock Phase transition in the number partitioning problem.
\newblock {\em Phys.\ Rev.\ Lett.}, 81(20):4281--4284, November 1998.

\bibitem{mertens:99a}
Stephan Mertens.
\newblock A complete anytime algorithm for balanced number partitioning.
\newblock preprint xxx.lanl.gov/abs/cs.DS/9903011, 1999.

\bibitem{mertens:00a}
Stephan Mertens.
\newblock Random costs in combinatorial optimization.
\newblock {\em Phys.\ Rev.\ Lett.}, 84(7):1347--1350, February 2000.

\bibitem{mezard:etal:87}
M.~M\'ezard, G.~Parisi, and M.A. Virasoro.
\newblock {\em Spin glass theory and beyond}.
\newblock World Scientific, Singapore, 1987.

\bibitem{provost:vallee:83}
J.P. Provost and G.~Vallee.
\newblock Ergodicity of the coupling constants and the symmetric $n$-replicas
  trick for a class of mean-field spin-glass models.
\newblock {\em Phys.\ Rev.\ Lett.}, 50(8):598--600, February 1983.

\bibitem{rieger:98}
Heiko Rieger.
\newblock Frustrated systems: Ground state properties via combinatorial
  optimization.
\newblock In J.~Kertesz and I.~Kondor, editors, {\em Advances in Computer
  Simulation}, volume 501 of {\em Lecture Notes in Physics}, Ber\-lin
  Hei\-del\-berg New~York, 1998. Spring\-er-Verlag.

\bibitem{monasson:etal:99}
R.Monasson, R.~Zecchina, S.~Kirkpatrick, B.~Selman, and L.~Troyanksy.
\newblock Determining computational complexity from characteristic `phase
  transitions'.
\newblock {\em Nature}, 400:133--137, July 1999.

\bibitem{ruml:etal:96}
W.~Ruml, J.T. Ngo, J.~Marks, and S.M. Shieber.
\newblock Easily searched encodings for number partitioning.
\newblock {\em Journal of Optimization Theory and Applications},
  89(2):251--291, May 1996.

\bibitem{tsai:92}
Li-Hui Tsai.
\newblock Asymptotic analysis of an algorithm for balanced parallel processor
  scheduling.
\newblock {\em {SIAM} J.\ Comput.}, 21(1):59--64, 1992.

\end{thebibliography}

\end{document}